\documentclass[aps,twocolumn,letterpaper,superscriptaddress]{revtex4-1}

\usepackage{amsmath}
\usepackage{amssymb}
\usepackage{amsfonts}
\usepackage{bm}
\usepackage{mathrsfs}
\usepackage{tensor}
\usepackage{dsfont}
\usepackage{esint}
\usepackage{comment}

\usepackage[usenames,dvipsnames]{xcolor}
\usepackage[hyperindex,pdftex, breaklinks,colorlinks = true,linkcolor = blue,urlcolor=blue,citecolor=blue]{hyperref}

\usepackage{physics}
\usepackage[section]{placeins}

\begin{document}
	
	\title{Linear response of a Chern insulator to finite-frequency electric fields}

	\author{Jason G. Kattan}
	\email{jkattan@physics.utoronto.ca}
	\affiliation{Department of Physics, University of Toronto, Toronto, Ontario M5S 1A7, Canada}

        \author{Alistair H. Duff}
	\email{aduff@physics.utoronto.ca}
	\affiliation{Department of Physics, University of Toronto, Toronto, Ontario M5S 1A7, Canada}
 
	\author{J. E. Sipe}
	\email{sipe@physics.utoronto.ca}
	\affiliation{Department of Physics, University of Toronto, Toronto, Ontario M5S 1A7, Canada}
	
	\date{\today }
	
	\begin{abstract}
             We derive the macroscopic charge and current densities of a Chern insulator initially occupying its electronic ground state as it responds to a finite-frequency electric field; we use a previously developed formalism based on microscopic polarization and magnetization fields in extended media. In a topologically trivial insulator, our result reduces to the familiar expression for the induced current density in linear response obtained from a Kubo analysis. But for a Chern insulator we find an extra ``topological" term involving the (first) Chern number associated with the occupied bands, encoding the quantum anomalous Hall effect in the presence of a frequency-dependent electric field. While an analogous term has been introduced in the ``modern theories of polarization and magnetization" for the linear response of finite-sized systems to static electric fields, our expression is valid for bulk Chern insulators in the presence of both static and finite-frequency electric fields, being derived analytically from a microscopic treatment of the electronic degrees of freedom, and can be generalized in a straightforward way to describe the response of a Chern insulator to electromagnetic fields that are not only frequency-dependent but also spatially inhomogeneous.	
        \end{abstract}
	
	\maketitle

\section{Introduction}\label{Sec1}

In the electrodynamics of material media, the classical Hall effect occurs when a conductor that is immersed in a static and spatially uniform magnetic field $\bm{B}$ is subject to an electric field $\bm{E}$ perpendicular to $\bm{B}$; a flow of charge carriers in the direction of $\bm{E} \times \bm{B}$ arises, resulting in a measurable potential difference across the boundary of the material \cite{Hall1, Klitzing1}. To linear order in the applied field, the macroscopic current is
\begin{equation}
    J^{i} = \sigma^{ij} E^j,
    \label{introJ}
\end{equation}
where $\sigma^{ij}$ are the components of the conductivity tensor, which depends implicitly on the background magnetic field; Latin indices denote Cartesian components. In general, this current density contains a part that is always perpendicular to the electric field, independent of how this field is oriented relative to the crystal axes. Indeed, the conductivity tensor in Eq. (\ref{introJ}) can be decomposed into its symmetric and antisymmetric parts, where the former characterizes the usual ``Ohmic current" parallel to the applied field, while the latter encodes the drift of charge carriers perpendicular to the applied field, often called the ``Hall current" \cite{Ashcroft}. The \textit{Hall conductivity} involves the off-diagonal components of the conductivity tensor, which in this classical analysis depend continuously on the magnitude of the magnetic field \cite{ClassicalHallexp1, ClassicalHallbook1}.

In quantum mechanics, however, the situation is different. The presence of a static and uniform magnetic field of magnitude $B$ along the $z$-direction results in the familiar Landau energy levels $E_n$ associated with quantized cyclotron orbits of charge carriers in the $xy$-plane \cite{Landaulevels, QAHeffect1}. Applying a static electric field of magnitude $E$ along the $x$-direction leads to modified Landau levels $E_{n k_y}$ depending on a momentum $\hbar k_y$, which causes a drift of charge carriers in the $y$-direction with group velocity \cite{QAHeffect2, QAHeffect3}
\begin{equation}
    \frac{1}{\hbar} \frac{\partial E_{n k_y}}{\partial k_y} =  - \frac{cE}{B}.
    \label{groupvel}
\end{equation}
This collective drift of charge carriers can be encoded in a \textit{quantized} Hall conductivity \cite{Laughlin1} of the form
\begin{equation}
    \sigma^{yx} = \frac{e^2}{2\pi \hbar} \nu,
\end{equation}
depending on some $\nu \in \mathbb{Z}$ related implicitly to the background magnetic field. This ``quantization" of the Hall conductivity is aptly referred to as the \textit{quantum Hall effect}. Of course, this analysis breaks down in the limit that the background magnetic field is ``turned off," since in this treatment based on Landau levels the group velocity (\ref{groupvel}) of the charge carriers in the material is ill-defined in the limit $B \to 0$. 

Surprisingly, however, there exists a class of band insulators for which an applied electric field leads to a non-vanishing quantized Hall conductivity even in the absence of a background magnetic field; this effect is referred to as the \textit{quantum anomalous Hall effect} \cite{QAHExperiment1}. In these ``Chern insulators," time-reversal symmetry is spontaneously broken by the presence of a static magnetic field that is \textit{intrinsic} to the system, meaning the charge carriers are distributed in such a way that they collectively generate an \textit{internal} magnetic field, even in a system at zero temperature occupying its electronic ground state \cite{QAHTheory1}. The quantized Hall conductivity for a two-dimensional Chern insulator takes the form \cite{TKNN}
\begin{equation}
    \sigma^{yx} = \frac{e^2 }{2\pi \hbar} C_{\mathcal{V}},
\end{equation}
where $C_{\mathcal{V}}$ is an integer encoding topological properties of the occupied valence bands in the crystal \footnote{The notation here differs from our previous work \cite{CImanuscript1} in that this conductivity tensor represents the macroscopic current induced in the $y$-direction, while there we considered the current induced in the $x$-direction}. The quantum anomalous Hall effect was first observed experimentally in two-dimensional thin films comprised of magnetically doped $\mathrm{Bi}_2 \mathrm{Te}_3$ or $\mathrm{Sb}_2 \mathrm{Te}_3$ \cite{QAHExperiment2, QAHExperiment3}, and in three-dimensional structures formed from layers thereof \cite{QAHExperiment5}. Magnetic doping in these compounds breaks time-reversal symmetry, which is necessary for a crystalline insulator to have a non-vanishing Chern number. It has also been realized in the magnetic topological insulator $\mathrm{Mn Bi}_2 \mathrm{Te}_4$ \cite{QAHExperiment6}, and in Moiré materials formed from graphene \cite{QAHExperiment8} and transition-metal dichalcogenides \cite{QAHExperiment7}.

Continuum models for Chern insulators are usually formulated within the so-called ``modern theories of polarization and magnetization" \cite{KingSmith1993, Resta1994}. In the former, a finite-sized band insulator is adiabatically perturbed by application of a static and uniform electric field, thereby inducing a current density $\bm{J}$ associated with the redistribution of charge carriers in the material. Then a macroscopic notion of polarization $\bm{P}$ is extracted by integrating the relation $\bm{J} = d\bm{P}/dt$ along the closed loop in parameter space formed by this adiabatic perturbation; a similar procedure is used to define a macroscopic notion of orbital magnetization \cite{Resta2010}. While the original papers introducing these ``modern theories" focused on topologically trivial insulators, the expressions therein have since been extended to Chern insulators by means of thermodynamic arguments \cite{Niu2007, Vanderbilt2009}. However, these arguments are really only applicable to finite-sized systems. For example, while the expression for orbital magnetization in a Chern insulator includes a term involving the product of a chemical potential and the Berry curvature \cite{Resta2006}, in an \textit{infinite} crystal the location of a chemical potential in the bandgap is essentially arbitrary \cite{VanderbiltBook, Girvin}. This is unlike the situation in a finite-sized system, where the chemical potential is fixed by properties of the boundary \cite{Hatsugai1, Hatsugai2}. Moreover, in the ``modern theories" the adiabatic perturbations used to extract notions of polarization and orbital magnetization are restricted to static applied fields, and it is unclear how to extend these arguments to the optical regime.

Recently we have implemented a formalism based on \textit{microscopic} polarization and magnetization fields in extended media \cite{Swiecicki2014, Mahon2019} to study an unperturbed Chern insulator occupying its zero-temperature electronic ground state \cite{CImanuscript1}. In the simpler case of topologically trivial insulators, these microscopic fields are defined using a complete set of exponentially localized Wannier functions (ELWFs) formed from Bloch energy eigenfunctions associated with the occupied valence bands, with macroscopic polarization and magnetization fields constructed by an appropriate spatial averaging procedure. And since charge carriers can flow between lattice sites, we introduce microscopic free charge and current densities, modelled through a generalized lattice gauge theory in such a way that the continuity equation for the free charge and current densities is satisfied. But an extension of such a strategy is necessary for Chern insulators, since it is well-known that no such set of ELWFs formed from the occupied bands alone exists \cite{Brouder}. Nonetheless, one \textit{can} introduce a set of ELWFs for a Chern insulator provided they are formed from Bloch energy eigenfunctions associated with both the occupied bands \textit{and} a sufficient number of unoccupied bands so that the total Chern number associated with all the bands involved vanishes. With this modified set of ELWFs, microscopic polarization and magnetization fields can be defined in a manner analogous to those of a trivial insulator, and one can then extract macroscopic expressions for polarization and orbital magnetization in a bulk Chern insulator \cite{CImanuscript1}. The expression so derived for the polarization was in agreement with that of the ``modern theories," but our expression for the orbital magnetization included a new term associated with the topological structure of the occupied valence bands, replacing the chemical-potential term of the ``modern theories" described above.

In this work we turn to a perturbed Chern insulator, identifying its response to applied electromagnetic fields in the optical regime. Here we restrict ourselves to spatially uniform macroscopic electric fields, making the common ``long-wavelength approximation" and neglecting effects due to macroscopic magnetic fields. We derive an expression for the macroscopic current density $\bm{J}(\omega)$ to linear order in the applied electric field $\bm{E}(\omega)$, which takes a similar form to Eq. (\ref{introJ}) except that all of the quantities therein now depend on the frequency $\omega$ of the applied field. In particular, we obtain a conductivity tensor $\sigma^{ij}(\omega)$ that is a sum of two contributions: One is the analog of the frequency-dependent term that is obtained from a Kubo analysis of topologically trivial insulators \cite{Kubo1, VanderbiltBook}, while the other is a frequency-independent response coefficient, unique to Chern insulators, that encodes the quantum anomalous Hall effect in three dimensions for a crystal with an arbitrary lattice structure. While the results here are valid only for uniform electric fields in linear response, this work lays the foundation for a wider research program involving the response of Chern insulators to generic optical fields in both the linear and nonlinear regimes, and to the inclusion of electron-electron interactions; here we restrict ourselves to the independent particle approximation.

We begin in Sec. \ref{Sec2} by reviewing a microscopic formulation of the electronic degrees of freedom in an unperturbed Chern insulator in terms of second-quantized electron field operators, leading to a definition of the ``single-particle density matrix" (SPDM) that is central to calculations in our formalism. We also give a detailed description of the geometry and topology of \textcolor{blue}{a} band structure necessary to define Chern numbers for generic crystalline insulators. In Sec. \ref{Sec3A} we use the single-particle density matrix to introduce microscopic charge and current densities associated with lattice sites in the crystal, which are used to define the microscopic fields. Then in Sec. \ref{Sec3B} we implement a spatial averaging procedure to obtain the corresponding macroscopic fields. Through a perturbative expansion of the SPDM in powers of the applied field, we obtain leading-order expressions for these macroscopic quantities in terms of this field and the geometric objects introduced in Sec. \ref{Sec2}; these are packaged into the final expression for the linearly induced current density in Sec. \ref{Sec4}, with the calculational details relegated to the Appendices. And we conclude in Sec. \ref{Sec5} by discussing some mathematical properties of our expression for the conductivity tensor, and draw comparisons to existing expressions in the literature.

\section{Single-particle density matrix}\label{Sec2}

To begin, we consider a three-dimensional bulk insulator at zero temperature, described within the independent particle approximation and initially occupying its electronic ground state. We implement the frozen-ion approximation, whereby the ions in the crystal's lattice are fixed in space, and we also neglect the spin degree of freedom of the electrons. Working in the Heisenberg picture, the electronic degrees of freedom of the system in the absence of an electromagnetic field are encoded in an electron field operator $\hat{\psi}_0(\bm{x},t)$, the dynamical evolution of which is governed by the Heisenberg equation
\begin{equation}
    i\hbar \frac{\partial \hat{\psi}_0(\bm{x},t)}{\partial t} = \big[\hat{\psi}_0(\bm{x},t),\hat{H}_0\big]
\end{equation}
with the Hamiltonian
\begin{equation}
    \hat{H}_0 = \int \mathrm{d}\bm{x}\, \hat{\psi}_{0}^{\dagger}(\bm{x},t) \mathcal{H}_0(\bm{x}) \hat{\psi}_0(\bm{x},t),
    \label{unperturbedH}
\end{equation}
where we have defined
\begin{equation}
    \mathcal{H}_0(\bm{x}) \equiv \frac{1}{2m} \big(\bm{\mathfrak{p}}(\bm{x})\big)^2 + \mathrm{V}_{\Gamma}(\bm{x}).
\end{equation}
Here $\mathrm{V}_{\Gamma}(\bm{x})$ is the cell-periodic background potential generated by the ions in the lattice $\Gamma$ of the crystal, and we have defined \cite{Mahon2019}
\begin{equation}
    \bm{\mathfrak{p}}(\bm{x}) \equiv \frac{\hbar}{i}\bm{\nabla} - \frac{e}{c} \bm{a}_{\text{static}}(\bm{x}),
\end{equation}
where $\bm{a}_{\text{static}}(\bm{x})$ is an ``internal," cell-periodic vector potential, the presence of which breaks time-reversal symmetry. Implementing Bloch's theorem with the Hamiltonian (\ref{unperturbedH}) yields the spectral problem
\begin{equation}
    \hat{H}_0 \ket{\psi_{n\bm{k}}} = E_{n\bm{k}} \ket{\psi_{n\bm{k}}},
    \label{spectralH0}
\end{equation}
where $n \in \mathbb{N}$ is a band index and $\bm{k}$ is a point in the (first) Brillouin zone, which is a smooth manifold $\text{BZ}$ that is diffeomorphic to the $3$-torus $\mathbb{T}^3$ \cite{Panati2017}. A characteristic feature of bulk insulators is that there exists an energy gap separating the first $N \in \mathbb{N}$ occupied ``valence" bands from the remaining unoccupied ``conduction" bands. The Bloch states $\ket{\psi_{n\bm{k}}}$ and their eigenvalues $E_{n\bm{k}}$ represent the valence bands when $1 \leq n \leq N$ and the conduction bands for all $n > N$. For each $\bm{k} \in \mathrm{BZ}$, the Bloch energy eigenfunctions $\psi_{n\bm{k}}(\bm{x}) \equiv \bra{\bm{x}}\ket{\psi_{n\bm{k}}}$ can be written \cite{Ashcroft}
\begin{equation}
    \psi_{n\bm{k}}(\bm{x}) = \frac{1}{(2\pi)^{3/2}} e^{i\bm{k}\cdot\bm{x}} u_{n\bm{k}}(\bm{x}),
\end{equation}
where the wavefunctions $u_{n\bm{k}}(\bm{x}) \equiv \bra{\bm{x}}\ket{n\bm{k}}$ are the cell-periodic parts thereof, and we refer to the states $\ket{n\bm{k}}$ as \textit{cell-periodic Bloch states}. While these cell-periodic Bloch states are defined canonically through Bloch's theorem, the physical quantities of interest to us may be expressed in any number of $\bm{k}$-dependent orthonormal (Schauder) bases obtained from unitary transformations of the form
\begin{equation}
    \ket{\alpha\bm{k}} = \sum_{n} U_{n\alpha}(\bm{k}) \ket{n\bm{k}},
    \label{unitarytransform1}
\end{equation}
where $U_{n\alpha}(\bm{k})$ are the components of a $\bm{k}$-dependent unitary operator $U(\bm{k})$. An important example of such quantities are \textit{exponentially localized Wannier functions} (ELWFs), defined in general by \cite{Brouder}
\begin{equation}
    W_{\alpha\bm{R}}(\bm{x}) \equiv \sqrt{\Omega_{\text{uc}}} \int_{\mathrm{BZ}} \frac{\mathrm{d}\bm{k}}{(2\pi)^3}\, e^{i\bm{k}\cdot(\bm{x}-\bm{R})} u_{\alpha\bm{k}}(\bm{x}),
    \label{unperturbedELWF}
\end{equation}
where $u_{\alpha\bm{k}}(\bm{x}) \equiv \bra{\bm{x}}\ket{\alpha\bm{k}}$ are the wavefunctions obtained from Eq. (\ref{unitarytransform1}) and $\Omega_{\text{uc}}$ is the volume of the unit cell. The specification of a complete set of ELWFs clearly depends on the above choice of unitary matrices, thereby yielding a kind of ``gauge freedom" that is distinct from the gauge freedom inherent in the $\mathrm{U}(1)$--gauge potentials encoding the applied electromagnetic field. And in constructing microscopic polarization and magnetization fields, which have contributions associated with each of the lattice sites in the crystal, these ELWFs will play a central role. Different choices for such a set of ELWFs will lead to different expressions for some of the material quantities below, although it is important that any \textit{physically measurable} quantities be independent of the kind of ``gauge freedom" inherent in Eq. (\ref{unitarytransform1}).

A natural strategy would be to construct one set of ELWFs from the valence bands and another set from the conduction bands, separately. However, it is well-known that there exist topological obstructions to constructing such ELWFs in certain classes of condensed matter systems, including Chern insulators \cite{Monaco2017}. To define the invariants that characterize these obstructions, we assemble the collection of cell-periodic Bloch states $\{\ket{n\bm{k}}\}_{n\in \mathbb{N}}$ for all $\bm{k}\in\mathrm{BZ}$ into a geometric structure called the \textit{Bloch bundle} \cite{Panati2017}, which is a smooth Hilbert bundle $\mathcal{B}$ wherein the fibre above a given $\bm{k} \in \mathrm{BZ}$ is the Hilbert space $\mathcal{B}_{\bm{k}}$ spanned by the cell-periodic Bloch states $\ket{n\bm{k}}$ for all $n \in \mathbb{N}$. In a bulk insulator for which there is a well-defined bandgap separating the occupied and unoccupied bands, the Bloch bundle admits a Whitney-sum decomposition $\mathcal{B} = \mathcal{V} \oplus \mathcal{C}$ into two subbundles $\mathcal{V}$ and $\mathcal{C}$ over the $\mathrm{BZ}$, where the fibres of the \textit{valence bundle} $\mathcal{V}$ are spanned pointwise by the set $\{\ket{n\bm{k}}\}_{n=1}^N$ of cell-periodic Bloch states representing the $N$ occupied valence bands, while the fibres of the \textit{conduction bundle} $\mathcal{C}$ are spanned pointwise by the remaining cell-periodic Bloch states representing the unoccupied conduction bands \cite{Fruchart1}. 

For a two-dimensional system, the relevant topological invariants are the first Chern numbers $C_{\mathcal{B}} = C_{\mathcal{V}} + C_{\mathcal{C}}$ of the Bloch bundle $\mathcal{B}$ and its valence and conduction subbundles $\mathcal{V}$ and $\mathcal{C}$ \cite{Monaco2017}, while in three dimensions there are a triple of such Chern numbers $C_{\mathcal{B}}^{\alpha} = C_{\mathcal{V}}^{\alpha} + C_{\mathcal{C}}^{\alpha}$ with $1 \leq \alpha \leq 3$ \cite{Troyer2016, Panati2017}. One can construct a Bloch bundle $\mathcal{B}$ from the $N$ occupied bands \textit{and} sufficiently many unoccupied bands such that its Chern number(s) vanish, implying that $\mathcal{B}$ is trivializable as a smooth Hilbert bundle. Notably, it has been shown that such a set always exists \cite{Freed}, although in general one must include all of the unoccupied bands; we take that general approach here. 

The Chern number(s) of a Hilbert bundle over the Brillouin zone encode the way in which its fibres ``twist" as the states therein are parallel transported between neighboring fibres by means of a connection. To make this precise, consider the \textit{frame bundle} associated to the Bloch bundle, the sections of which are $\bm{k}$-dependent orthonormal Schauder bases (or \textit{frames}) over open sets in the Brillouin zone, an example of which is the locally-defined ``Bloch frame" $(\ket{n \bm{k}})_{n \in \mathbb{N}}$. Since the Bloch bundle is trivializable by construction, its frame bundle admits global sections --- that is, globally-defined frames for the Bloch bundle --- which we will refer to as \textit{Wannier frames} $(\ket{\alpha\bm{k}})_{\alpha \in \mathbb{N}}$. As this notation suggests, the elements of a Wannier frame are obtained from the cell-periodic Bloch states by unitary transformations of the form (\ref{unitarytransform1}). If we equip the frame bundle with a connection, which induces a connection on the Bloch bundle, then \textit{local} frames for the Bloch bundle yield \textit{local} component representations of this connection \cite{HamiltonBook}. There is a natural choice of connection for bulk crystalline insulators, called the \textit{non-Abelian Berry connection} (see Appendix \ref{AppendixA}), and its local component representation in the Bloch frame $(\ket{n\bm{k}})_{n\in \mathbb{N}}$ is given by
\begin{align}
    \xi_{mn}^i(\bm{k}) = i \big(m\bm{k}|\partial_i n\bm{k}\big),\label{BerryconnectionBloch}
\end{align}
where $\partial_i \equiv \partial / \partial k^i$ and the inner product on the Hilbert space spanned by the cell-periodic parts of the Bloch energy eigenfunctions is defined by
\begin{equation}
    (f|g) \equiv \frac{1}{\Omega_{\text{uc}}} \int_{\Omega} \mathrm{d}\bm{x}\, f^*(\bm{x}) g(\bm{x}).
    \label{innerproduct}
\end{equation}

There is an important subtlety in the expression (\ref{BerryconnectionBloch}) related to band crossings throughout the Brillouin zone. Denote by $\mathscr{D}_{\mathcal{B}} \subseteq \mathrm{BZ}$ the locus of degeneracies of the band structure encoded in the Bloch bundle; that is, the set of points $\bm{k}_0 \in \mathrm{BZ}$ for which $E_{n\bm{k}_0} = E_{m\bm{k}_0}$ for some pair of bands indexed by $n,m \in \mathbb{N}$. While a Wannier frame is globally smooth by definition, the Bloch frame $(\ket{n\bm{k}})_{n\in \mathbb{N}}$ is not smooth at points $\bm{k}_0 \in \mathscr{D}_{\mathcal{B}}$ and the local component representation (\ref{BerryconnectionBloch}) becomes singular, so this expression is to be understood only for those points of the Brillouin zone not contained in $\mathscr{D}_{\mathcal{B}}$. But since the elements of a Wannier frame $(\ket{\alpha\bm{k}})_{\alpha \in \mathbb{N}}$ \textit{are} smooth across the Brillouin zone, the \textit{global} component representation of the Berry connection, given by
\begin{equation}
    \tilde{\xi}_{\alpha\beta}^i(\bm{k}) = i \big(\alpha\bm{k}|\partial_i \beta\bm{k}\big),
    \label{BerryconnectionWannier}
\end{equation}
is well-defined for all points in the Brillouin zone. Recalling that elements of the Bloch and Wannier frames are related by Eq. (\ref{unitarytransform1}), the corresponding component representations of the Berry connection are related locally by the unitary transformation
\begin{equation}
    \sum_{\alpha\beta} U_{m\beta}(\bm{k}) \tilde{\xi}_{\beta\alpha}^i(\bm{k}) U_{\alpha n}^{\dagger}(\bm{k}) = \xi_{mn}^i(\bm{k}) + \mathcal{W}_{mn}^i(\bm{k}),
    \label{BlochWannierconnection}
\end{equation}
where we have introduced the Hermitian operator $\mathcal{W}^i(\bm{k})$ populated by the Bloch-frame matrix elements
\begin{equation}
    \mathcal{W}_{mn}^i(\bm{k}) \equiv i \sum_{\alpha} \big(\partial_i U_{m\alpha}(\bm{k})\big) U_{\alpha n}^{\dagger}(\bm{k}).\label{Wmatrix}
\end{equation}

Like the local component representation (\ref{BerryconnectionBloch}) of the Berry connection, the two expressions (\ref{BlochWannierconnection}) and (\ref{Wmatrix}) are also only well-defined at points $\bm{k} \in \mathrm{BZ}$ not contained in the set $\mathscr{D}_{\mathcal{B}}$. Even where they are well-defined, there is a ``gauge-dependence" of these component representations on the choice of Wannier frame, which is encoded in the quantity $\mathcal{W}_{mn}^i(\bm{k})$ that is related to the Maurer-Cartan form for the structure group $\mathrm{U}(\mathcal{B})$ of the frame bundle. In terms of the components of the Berry connection in Eq. (\ref{BerryconnectionBloch}), the first Chern number of the valence bundle is given in two dimensions by
\begin{equation}
    C_{\mathcal{V}} = \frac{1}{2\pi} \sum_{n} f_n \fint_{\mathrm{BZ}} \mathrm{d}\bm{k} \left(\frac{\partial \xi_{nn}^y(\bm{k})}{\partial k_x} - \frac{\partial \xi_{nn}^x(\bm{k})}{\partial k_y}\right),
    \label{Chernnumber2d}
\end{equation}
where $f_n$ is the filling factor for the $n\text{th}$ band ($f_n=1$ for valence bands and $f_n=0$ for conduction bands). Here we have introduced an ``adapted integral" over the $\mathrm{BZ}$, denoted by a bar through the integral symbol, the definition of which is given in Appendix \ref{AppendixA4} and is related to subtleties involving band crossings for points in $\mathscr{D}_{\mathcal{B}}$. A two-dimensional Chern insulator is a crystalline insulator for which the Chern number $C_{\mathcal{V}}$ is nonzero in its unperturbed ground state, which requires broken time-reversal symmetry. Meanwhile, in three dimensions the (Berry) curvature associated with the component representation (\ref{BerryconnectionBloch}) of the Berry connection in the Bloch frame is \cite{NakaharaBook} 
\begin{equation}
    F_{mn}^{ij}(\bm{k}) = \frac{\partial \xi_{mn}^j(\bm{k})}{\partial k^i} - \frac{\partial \xi_{mn}^i(\bm{k})}{\partial k^j} - i \big[\xi^i(\bm{k}), \xi^{j}(\bm{k})\big]_{mn},
    \label{Berrycurvaturecomponents}
\end{equation}
where $\bm{k} \in \mathrm{BZ} \setminus \mathscr{D}_{\mathcal{B}}$ and the matrix elements of the Lie bracket are
\begin{equation}
    \big[\xi^i(\bm{k}),\xi^j(\bm{k})\big]_{mn} \equiv \sum_{\ell}\Big(\xi_{m\ell}^i(\bm{k}) \xi_{\ell n}^j(\bm{k}) - \xi_{m \ell}^j(\bm{k}) \xi_{\ell n }^i(\bm{k})\Big).
\end{equation}
The Chern number $C_{\mathcal{V}}^{\alpha}$ is the trace over the occupied bands of this Berry curvature integrated over the $\alpha\text{th}$ $2$-cycle of the Brillouin zone $\text{BZ} \cong \mathbb{T}^3$, and a three-dimensional Chern insulator is a crystalline insulator for which \textit{at least one} of the Chern numbers $C_{\mathcal{V}}^{\alpha}$ is nonzero in its electronic ground state, again requiring broken time-reversal symmetry. In both two and three dimensions, if there is no nonzero Chern number, then we refer to the system as a (\textit{topologically}) \textit{trivial insulator}. 

For a Chern insulator in three dimensions, at least one of the Chern numbers $C_{\mathcal{V}}^{\alpha}$ for the valence bundle $\mathcal{V}$ is nonzero. These Chern numbers constitute a topological obstruction to finding a set of ``quasi-Bloch functions" $u_{\alpha\bm{k}}(\bm{x}) \equiv \bra{\bm{x}}\ket{\alpha\bm{k}}$ that are globally analytic across the $\mathrm{BZ}$ \cite{Brouder}, using the unitary transformation (\ref{unitarytransform1}) with the sum restricted to the occupied valence bands ($1 \leq n \leq N$). Since the Wannier functions (\ref{unperturbedELWF}) are only exponentially localized if these quasi-Bloch functions are globally analytic, it follows that the nonvanishing of these Chern numbers constitutes a topological obstruction to the construction of ELWFs through Eq. (\ref{unperturbedELWF}) from the occupied valence bands alone.

However, the entire Bloch bundle $\mathcal{B}$, which is formed from the occupied valence bands \textit{and} all of the unoccupied conduction bands, is trivializable, and so its Chern numbers $C_{\mathcal{B}}^{\alpha} = 0$ for all $1 \leq \alpha \leq 3$ \cite{Panati2017}. So then it \textit{is} possible to construct a set of quasi-Bloch functions that are globally analytic across the $\mathrm{BZ}$, provided we enlarge the sum in Eq. (\ref{unitarytransform1}) to include all of the bands that are collectively described by the Bloch bundle; since the Chern numbers for the Bloch bundle vanish, it is always possible to find such a set. And from this set we can construct a set of ELWFs for a Chern insulator, with the only difference being that they are constructed from cell-periodic Bloch states associated with all of the bands in general \cite{CImanuscript1}. Writing the ELWFs in the form $W_{\alpha\bm{R}}(\bm{x}) = \bra{\bm{x}}\ket{\alpha\bm{R}}$, the states $\ket{\alpha\bm{R}}$ are obtained from a given Wannier frame $(\ket{\alpha\bm{k}})_{\alpha \in \mathbb{N}}$ by the Bloch-Floquet transform \cite{Panati2017}
\begin{equation}
    \bra{\bm{x}}\ket{\alpha\bm{R}} = \sqrt{\Omega_{\text{uc}}} \int_{\text{BZ}} \frac{\mathrm{d}\bm{k}}{(2\pi)^3}\, e^{i\bm{k}\cdot(\bm{x}-\bm{R})} \bra{\bm{x}}\ket{\alpha\bm{k}}.
\end{equation}
Then introducing a collection of creation and annihilation operators such that $\ket{\alpha\bm{R}} = \hat{a}_{\alpha\bm{R}}^{\dagger} \ket{\text{vac}}$ for all $\alpha \in \mathbb{N}$, the \textit{single-particle density matrix} (SPDM) for the unperturbed system is defined by
\begin{equation}
    \eta_{\alpha\bm{R}'';\beta\bm{R}'}^{(0)} \equiv \bra{\text{gs}} \hat{a}_{\beta\bm{R}'}^{\dagger} \hat{a}_{\alpha\bm{R}''}\ket{\text{gs}},
    \label{unperturbedSPDM}
\end{equation}
where $\ket{\text{gs}}$ is the electronic ground state of the system. We have shown previously \cite{CImanuscript1} that the SPDM for an unperturbed Chern insulator is given by
\begin{align}
    \eta_{\alpha\bm{R}'';\beta\bm{R}'}^{(0)} = \frac{\Omega_{\text{uc}}}{(2\pi)^3}\sum_n f_n \fint_{\mathrm{BZ}} \mathrm{d}\bm{k}\, e^{i\bm{k}\cdot(\bm{R}'' - \bm{R}')}U_{\alpha n}^{\dagger}(\bm{k}) U_{n\beta}(\bm{k}).
\end{align}
In the limit of a trivial insulator, the states $\ket{\alpha\bm{k}}$ can be constructed from superpositions of either the occupied or the unoccupied cell-periodic Bloch states alone, thereby rendering $U_{n\alpha}(\bm{k})$ block diagonal with ``upper block" of size $N \times N$ associated with the valence bands. Introducing a Fermi factor $f_{\alpha}$ for the states $\ket{\alpha\bm{k}}$, the SPDM for an unperturbed trivial insulator becomes
\begin{align}
    \eta_{\alpha\bm{R}'';\beta\bm{R}'}^{(0;\text{trivial})} = f_{\alpha} \delta_{\alpha\beta} \delta_{\bm{R}''\bm{R}'}.
\end{align}

For both trivial and nontrivial insulators in the independent particle approximation --- and with the neglect of local field corrections --- we can treat the interaction between the electrons and the Maxwell field with the usual minimal coupling prescription in the Hamiltonian (\ref{unperturbedH}). In the presence of an external vector potential $\bm{A}(\bm{x},t)$ describing the Maxwell field, the ELWFs discussed above must be adjusted by the inclusion of a generalized Peierls phase \cite{Mahon2019}
\begin{equation}
    \Phi(\bm{x},\bm{R};t) \equiv \frac{e}{\hbar c} \int \mathrm{d}\bm{y}\, s^i(\bm{y};\bm{x},\bm{R}) A^i(\bm{y},t),
\end{equation}
leading to the adjusted ELWFs
\begin{equation}
    W_{\alpha\bm{R}}'(\bm{x},t) = e^{i\Phi(\bm{x},\bm{R};t)} W_{\alpha\bm{R}}(\bm{x}).
\end{equation}
\\
Here we have introduced a ``relator" \cite{Mahon2019}
\begin{equation}
    \bm{s}(\bm{y};\bm{x},\bm{R}) = \int_{C(\bm{x},\bm{R})} \mathrm{d}\bm{z}\, \delta(\bm{y} - \bm{z}),
    \label{relatorS}
\end{equation}
where $\bm{z}: \mathbb{R} \to \mathbb{R}^3$ is a smooth curve with image path $C(\bm{x},\bm{R})$ that begins at a lattice site $\bm{R} \in \Gamma$ and ends at a point $\bm{x} \in \mathbb{R}^3$. Since these adjusted ELWFs no longer form an orthonormal set, we apply Löwdin's method \cite{Mayer} for symmetric orthogonalization and thereby obtain a suitable collection of modified ELWFs $\bar{W}_{\alpha\bm{R}}(\bm{x},t)$ with which to construct quantities for the perturbed system.

To account for the applied electromagnetic field in the single-particle density matrix of the interacting system, we introduce a perturbative expansion \cite{Mahon2020, Mahon2020a}
\begin{equation}
    \eta_{\alpha\bm{R}'';\beta\bm{R}'}(t) = \eta_{\alpha\bm{R}'';\beta\bm{R}'}^{(0)} + \eta_{\alpha\bm{R}'';\beta\bm{R}'}^{(1)}(t) + \dots,\label{expansionSPDM}
\end{equation}
where the superscript ``$(0)$" indicates the contribution that is independent of the applied field, the superscript ``$(1)$" denotes the contribution that is linear in the applied field, and so on. The zeroth-order SPDM in this expansion is equivalent to that of the unperturbed system given in Eq. (\ref{unperturbedSPDM}). Restricting to spatially homogeneous electric fields $\bm{E}(t)$ in the long-wavelength approximation, and implementing a Fourier-series decomposition
\begin{equation}
    g(t) = \sum_{\omega} e^{-i\omega t} g(\omega),\label{Fourieranalysis}
\end{equation}
we show in Appendix \ref{AppendixB} that the first-order contribution to the single-particle density matrix is given by
\begin{widetext}
\begin{equation}
    \eta_{\alpha\bm{R}'';\beta\bm{R}'}^{(1)}(\omega) = e  E^{\ell}(\omega) \frac{\Omega_{\text{uc}}}{(2\pi)^3} \sum_{mn} f_{nm} \fint_{\text{BZ}} \mathrm{d}\bm{k}\, e^{i\bm{k}\cdot(\bm{R}''-\bm{R}')} \frac{U_{\alpha m}^{\dagger}(\bm{k}) \xi_{mn}^{\ell}(\bm{k}) U_{n\beta}(\bm{k})}{E_{m\bm{k}} - E_{n\bm{k}} - \hbar(\omega + i 0^+)},
    \label{firstorderSPDMWannier}
\end{equation}
\end{widetext}
where $f_{nm} \equiv f_n - f_m$. It should be noted that this expression holds for both trivial and nontrivial insulators, but we show in the next section that there are important modifications to the induced current density in a Chern insulator.

\section{Polarization and magnetization}\label{Sec3}
Given microscopic expressions for the charge and current densities $(\rho,\bm{j})$ in an extended system, our approach is to introduce \textit{microscopic} polarization and magnetization fields $(\bm{p},\bm{m})$, together with free charge and current densities $(\rho_F,\bm{j}_F)$. In the first subsection below we sketch the general framework for obtaining these microscopic quantities in the presence of arbitrary electromagnetic fields; in the second subsection we introduce the corresponding macroscopic quantities in the limit of a uniform macroscopic electric field.

\subsection{Microscopic quantities}\label{Sec3A}
The microscopic polarization and magnetization fields, and the free charge and current densities, are related to the microscopic charge and current densities by \cite{Healybook}
\begin{align}
    \rho(\bm{x},t) &= - \bm{\nabla} \cdot \bm{p}(\bm{x},t) + \rho_F(\bm{x},t),\nonumber \\
    \bm{j}(\bm{x},t) &= \frac{\partial \bm{p}(\bm{x},t)}{\partial t} + c \bm{\nabla} \times \bm{m}(\bm{x},t) + \bm{j}_F(\bm{x},t).
    \label{microscopicpolmag}
\end{align}
This kind of formalism, based on microscopic polarization and magnetization fields, is standard in the theory of optical response for atoms and molecules \cite{Healybook, Kattan1}. In an extended system, we define the microscopic charge density $\rho(\bm{x},t) \equiv \langle\hat{\rho}(\bm{x},t)\rangle + \rho^{\text{ion}}(\bm{x})$ and the current density $\bm{j}(\bm{x},t) \equiv \langle\hat{\bm{j}}(\bm{x},t)\rangle$, involving expectation values of the electronic charge and current density operators. Here the charge density of point-like ions at each site $\bm{R}$ is
\begin{equation}
    \rho_{\bm{R}}^{\text{ion}}(\bm{x}) = \sum_{N} q_N \delta(\bm{x} - \bm{R} - \bm{d}_N),
\end{equation}
where $q_N$ denotes the charge of the $N\text{th}$ ion located at $\bm{R} + \bm{d}_N$ in the unit cell at $\bm{R}$, and the total ionic charge density is the sum over $\bm{R} \in \Gamma$ of this expression. 

Through a Green-function formalism introduced previously \cite{Mahon2019}, it has been shown that the expectation values of the electronic charge and current density operators admit a lattice-site decomposition
\begin{align}
    \langle\hat{\rho}(\bm{x},t)\rangle &= \sum_{\bm{R}} \rho_{\bm{R}}^{e}(\bm{x},t),\nonumber \\
    \langle\hat{\bm{j}}(\bm{x},t)\rangle &= \sum_{\bm{R}} \bm{j}_{\bm{R}}(\bm{x},t),
\end{align}
where the quantities on the right-hand-side are written in terms of the SPDM as
\begin{align}
    \rho_{\bm{R}}^{e}(\bm{x},t) &= \sum_{\alpha\beta\bm{R}'\bm{R}''} \rho_{\beta\bm{R}';\alpha\bm{R}''}(\bm{x},\bm{R};t) \eta_{\alpha\bm{R}'';\beta\bm{R}'}(t),\nonumber \\
    \bm{j}_{\bm{R}}(\bm{x},t) &= \sum_{\alpha\beta\bm{R}'\bm{R}''} \bm{j}_{\beta\bm{R}';\alpha\bm{R}''}(\bm{x},\bm{R};t) \eta_{\alpha\bm{R}'';\beta\bm{R}'}(t),
    \label{siterhoJ}
\end{align}
involving the so-called ``generalized site-quantity matrix elements" $\rho_{\beta\bm{R}';\alpha\bm{R}''}(\bm{x},\bm{R};t)$ and $\bm{j}_{\beta\bm{R}';\alpha\bm{R}''}(\bm{x},\bm{R};t)$, which are functions of the ELWFs and their derivatives \cite{Mahon2019}. A site polarization field $\bm{p}_{\bm{R}}(\bm{x},t)$ and a site magnetization field $\bm{m}_{\bm{R}}(\bm{x},t)$ can then be defined, with the total polarization and magnetization fields given by
\begin{align}
     \bm{p}(\bm{x},t) &= \sum_{\bm{R}} \bm{p}_{\bm{R}}(\bm{x},t),\nonumber \\
     \bm{m}(\bm{x},t) &= \sum_{\bm{R}} \bm{m}_{\bm{R}}(\bm{x},t),
\end{align}
and expressions for these quantities are given in Appendix \ref{AppendixC}. And as we discuss there, the site polarization and magnetization fields can be expanded in a series of generalized electric and magnetic multipole moments. In particular, the electric and magnetic dipole moments are 
\begin{align}
    \bm{\mu}_{\bm{R}}(t) &= \int \mathrm{d}\bm{x}\, \bm{p}_{\bm{R}}(\bm{x},t),\nonumber \\
    \bm{\nu}_{\bm{R}}(t) &= \int \mathrm{d}\bm{x}\, \bm{m}_{\bm{R}}(\bm{x},t).\label{dipolemoments}
\end{align}

In addition to the contributions from the site polarization and magnetization fields, there are also free charges and currents, and the microscopic free charge and current densities featured in Eq. (\ref{microscopicpolmag}) are modelled through a generalized lattice gauge theory. At each lattice site $\bm{R}$ we define the site charge
\begin{equation}
    Q_{\bm{R}}(t) \equiv \int \mathrm{d}\bm{x}\, \Big(\rho_{\bm{R}}^e(\bm{x},t) + \rho_{\bm{R}}^{\text{ion}}(\bm{x})\Big),
\end{equation}
in terms of which the free charge density is taken to be
\begin{equation}
    \rho_F(\bm{x},t) = \sum_{\bm{R}} Q_{\bm{R}}(t) \delta(\bm{x} - \bm{R}).\label{microscopicfreecharge}
\end{equation}
By examining the dynamical evolution of the site charges, one can show that the free current density is given by
\begin{equation}
    \bm{j}_F(\bm{x},t) = \frac{1}{2} \sum_{\bm{R}\bm{R}'} \bm{s}(\bm{x};\bm{R},\bm{R}') I(\bm{R},\bm{R}';t),
    \label{microscopicfreecurrent}
\end{equation}
where $I(\bm{R},\bm{R}';t)$ is the link current connecting site charges at the lattice sites $\bm{R}$ and $\bm{R}'$, which satisfies
\begin{equation}
    \frac{d Q_{\bm{R}}(t)}{dt} = \sum_{\bm{R}'} I(\bm{R},\bm{R}';t).
\end{equation}
The analytical expression for these link currents is given in Appendix \ref{AppendixC}.

\subsection{Macroscopic quantities}\label{Sec3B}
From the microscopic quantities defined above we obtain the corresponding macroscopic quantities through spatial averaging. Choosing an averaging function with a characteristic length scale $\Delta$ satisfying
\begin{equation}
    a \ll \Delta \ll \lambda,
\end{equation}
where $a$ is on the order of a lattice constant and $\lambda$ characterizes the typical range of variation of the Maxwell fields, the macroscopic polarization and magnetization fields are given by \cite{Mahon2020, Mahon2020a}
\begin{align}
    \bm{P}(\bm{x},t) &= \int \mathrm{d}\bm{y}\, w(\bm{x}-\bm{y}) \bm{p}(\bm{y},t),\nonumber \\
    \bm{M}(\bm{x},t) &= \int \mathrm{d}\bm{y}\, w(\bm{x}-\bm{y}) \bm{m}(\bm{y},t),\label{spatialaverage}
\end{align}
where $w(\bm{x}-\bm{y})$ is a smooth positive function that is spherically symmetric and drops off continuously as the distance $\|\bm{x}-\bm{y}\| \to \infty$ \cite{Mahon2020a}.

Consider first the ground-state values of the macroscopic polarization and magnetization, which we denote by $\bm{P}^{(0)}$ and $\bm{M}^{(0)}$ as they are zeroth-order in the response to the applied field. They will be independent of time and will only acquire contributions from the dipole moments $\bm{\mu}_{\bm{R}}$ and $\bm{\nu}_{\bm{R}}$ respectively, since the higher order multipole moments would lead only to contributions to $\bm{P}$ and $\bm{M}$ that would involve the variation of the macroscopic averages of those higher order moments; in the ground state those macroscopic averages are essentially uniform \cite{Mahon2020, Mahon2020a}. The zeroth-order expressions for the macroscopic polarization and magnetization fields were derived previously \cite{CImanuscript1}, and are given in Appendix \ref{AppendixC}. 

Turning to the first-order response to the electromagnetic field, in the long wavelength limit we need only keep the time-dependent contribution to the macroscopic polarization $\bm{P}$ that is uniform, and we need not consider the time-dependent response of the magnetization, since only the former will lead to a uniform time-dependent current density 
\cite{Mahon2020}; thus we can take
\begin{align}
    \bm{P}(t) &= \bm{P}^{(0)} + \bm{P}^{(1)}(t) + \dots,\nonumber \\
    \bm{M}(t) &= \bm{M}^{(0)} +  \dots,
\end{align}
explicitly including the unperturbed polarization and magnetization together with the only additional part of the fields $\bm{P}(t)$ and $\bm{M}(t)$ that will contribute to linear response in the long wavelength limit. With the usual expansion \cite{Mahon2020a}, we show in Appendix \ref{AppendixC} that the first-order contribution to the macroscopic polarization is given by
\begin{widetext}
\begin{align}
    P^{(1)i}(\omega) &= e^2 E^{\ell}(\omega) \sum_{mn} f_{nm} \fint_{\text{BZ}} \frac{\mathrm{d}\bm{k}}{(2\pi)^3}\, \frac{\big(\xi_{nm}^i(\bm{k}) + \mathcal{W}_{nm}^i(\bm{k})\big)\xi_{mn}^{\ell}(\bm{k})}{E_{m\bm{k}} - E_{n\bm{k}} - \hbar(\omega + i0^+)}.
    \label{macroscopicpolarization2}
\end{align}
\end{widetext}

For the macroscopic free charge density in Eq. (\ref{microscopicfreecharge}), we implement the averaging procedure above, leading to
\begin{equation}
    \varrho_F(t) = \frac{e}{\Omega_{\text{uc}}} \sum_{\alpha} \eta_{\alpha\bm{R};\alpha\bm{R}}(t)
\end{equation}
for any fixed $\bm{R} \in \Gamma$. With the perturbative expansion (\ref{expansionSPDM}) for the single-particle density matrix, the zeroth-order and first-order contributions to this can be shown to vanish for a system that is electrically neutral. Similarly, implementing our spatial averaging procedure in Eq. (\ref{microscopicfreecurrent}) and making the long-wavelength approximation, we find for the macroscopic free current density
\begin{equation}
    J_F^i(t) = \frac{1}{2\Omega_{\text{uc}}} \sum_{\bm{R}'} \big(R^i - R'^i\big) I(\bm{R},\bm{R}';t),
\end{equation}
for any fixed $\bm{R} \in \Gamma$. As discussed in Appendix \ref{AppendixC}, from a perturbative expansion of the link currents
\begin{equation}
    I(\bm{R},\bm{R}';t) = I^{(0)}(\bm{R},\bm{R}') + I^{(1)}(\bm{R},\bm{R}';t) + \dots,
    \label{linkexpansion}
\end{equation}
it has been shown that the zeroth-order contribution vanishes \cite{Mahon2019}. And the first-order contribution is 
\begin{widetext}
\begin{align}
    J_{F}^{(1)i}(\omega) = &+ i \omega e^2 E^{\ell}(\omega) \sum_{mn} f_{nm} \fint_{\text{BZ}} \frac{\mathrm{d}\bm{k}}{(2\pi)^3}\,  \frac{ \mathcal{W}_{nm}^i(\bm{k}) \xi_{mn}^{\ell}(\bm{k})}{E_{m\bm{k}} - E_{n\bm{k}} - \hbar (\omega + i0^+)}\nonumber \\
    &- \frac{e^2}{\hbar} E^{\ell}(\omega) \sum_{mn} f_{nm} \fint_{\text{BZ}} \frac{\mathrm{d}\bm{k}}{(2\pi)^3}\, \Im\big[\mathcal{W}_{nm}^{\ell}(\bm{k})\mathcal{W}_{mn}^i(\bm{k})\big].
    \label{macroscopicfreecurrent2}
\end{align}
\end{widetext}
Returning to the first-order polarization (\ref{macroscopicpolarization2}), notice that the first term in the integrand there is the usual term for the first-order modification of the macroscopic polarization in a trivial insulator, while the second term is novel to Chern insulators. While this novel term cancels the first line of the free current density (\ref{macroscopicfreecurrent2}) in the expression for the full current density, as we will see below, it cannot be moved to the free current through a gauge transformation between Wannier frames without modifying the form of the other terms therein; the same kind of split between the first-order polarization and free current occurs in the linear response of metals \cite{Mahon2021}.

\section{Macroscopic current density}\label{Sec4}

As discussed in Sec. \ref{Sec2}, any physically relevant quantities should be independent of the choice of ``gauge," that is, the choice of Wannier frame used to define a complete set of exponentially-localized Wannier functions through Eq. (\ref{unperturbedELWF}). It is straightforward to show that the first-order contributions to the macroscopic polarization (\ref{macroscopicpolarization2}) and free current (\ref{macroscopicfreecurrent2}) are ``gauge-dependent," owing to the presence of the matrix elements $\mathcal{W}_{mn}^i(\bm{k})$. However, we have argued previously that the macroscopic polarization and magnetization fields in an extended system are not physically measurable quantities. Instead, the macroscopic charge and current densities, obtained through the macroscopic relations \cite{Jackson}
\begin{align}
    \varrho(\bm{x},t) &= - \bm{\nabla} \cdot \bm{P}(\bm{x},t) + \varrho_{F}(\bm{x},t),\nonumber \\
    \bm{J}(\bm{x},t) &= \frac{\partial \bm{P}(\bm{x},t)}{\partial t} + c \bm{\nabla}\times\bm{M}(\bm{x},t) + \bm{J}_F(\bm{x},t),
    \label{macroscopicpolmag}
\end{align}
\textit{are} physically measurable quantities and, indeed, are ``gauge-invariant" in the sense described above. Making the long-wavelength approximation, in which all of these quantities become spatially homogeneous, implies that the macroscopic charge density is just the free charge density $\varrho_F(t)$, while the macroscopic current density is given by
\begin{align}
    \bm{J}(t) = \frac{d \bm{P}(t)}{dt} + \bm{J}_F(t).\label{rhoJLWA}
\end{align}
Restricting to linear response in the presence of a spatially uniform electric field, the macroscopic free charge density and hence the total charge density vanishes. Of course, this will no longer be true when a spatially inhomogeneous electromagnetic field is applied, since then the macroscopic polarization field will depend on position and may contribute to the first line of Eq. (\ref{macroscopicpolmag}). Moreover, since the zeroth-order link current $I^{(0)}(\bm{R},\bm{R}') = 0$, the zeroth-order contribution to the macroscopic free current density vanishes. Meanwhile, the first-order modification is
\begin{equation}
    \bm{J}^{(1)}(\omega) = - i \omega \bm{P}^{(1)}(\omega) + \bm{J}_F^{(1)}(\omega).
\end{equation}
\\
After combining Eqs. (\ref{macroscopicpolarization2}) and (\ref{macroscopicfreecurrent2}), it is straightforward to show that the first-order modification of the macroscopic current density is given by
\begin{widetext} 
\begin{align}
    J^{(1)i}(\omega) = - i \omega e^2 E^{\ell}(\omega) \sum_{mn} f_{nm} \fint_{\text{BZ}} \frac{\mathrm{d}\bm{k}}{(2\pi)^3}\, \frac{\xi_{nm}^i(\bm{k}) \xi_{mn}^{\ell}(\bm{k})}{E_{m\bm{k}} - E_{n\bm{k}} - \hbar(\omega + i0^+)} + \frac{e^2}{\hbar} \varepsilon^{i\ell j} E^{\ell}(\omega) \sum_{n} f_n \fint_{\text{BZ}} \frac{\mathrm{d}\bm{k}}{(2\pi)^3}\, \varepsilon^{j ab} \partial_a \mathcal{W}_{nn}^{b}(\bm{k}).
    \label{J1identity}
\end{align}
\end{widetext}
The first term takes the form of the usual frequency-dependent response of a trivial insulator, while the second term is sensitive to the topology of the valence bundle, or, more precisely, to the topological properties of the frame bundle that is associated to the valence bundle. Indeed, in Appendix \ref{AppendixD} we show that the second term depends on the (first) Chern numbers $C_{\mathcal{V}}^{\alpha}$ for all $1 \leq \alpha \leq 3$, which are obtained by integrating the Chern class of the valence bundle over the three fundamental $2$-cycles in the Brillouin zone; this term also depends on the properties of the crystal's lattice. In the static limit ($\omega \to 0^+$) the first term vanishes, and we obtain the usual quantum anomalous Hall contribution to the first-order induced current density in the presence of a static electric field. To differentiate these two contributions for finite frequencies, it is useful to write this expression as
\begin{equation}
    J^{(1)i}(\omega) = \sigma_{\text{Kubo}}^{i\ell}(\omega) E^{\ell}(\omega) + \sigma_{\text{QAH}}^{i\ell} E^{\ell}(\omega),
    \label{centralresult}
\end{equation}
where the first term,
\begin{widetext}
\begin{equation}
    \sigma_{\text{Kubo}}^{i\ell}(\omega) \equiv -i \omega e^2 \sum_{mn} f_{nm} \int_{\text{BZ}} \frac{\mathrm{d}\bm{k}}{(2\pi)^3}\, \frac{ \xi_{nm}^i(\bm{k})\xi_{mn}^{\ell}(\bm{k})}{E_{m\bm{k}} - E_{n\bm{k}} - \hbar(\omega + i0^+)},
    \label{Kuboconductivitytensor}
\end{equation}
\end{widetext}
is the usual Kubo form for the conductivity, and would be the only contribution were the insulator topologically trivial. Here we have dropped the ``adapted integral" notation because this expression vanishes at all points in the locus of degeneracies $\mathscr{D}_{\mathcal{B}}$ for which two valence (or conduction) bands intersect, owing to the presence of the factor $f_{nm} = f_n - f_m$. The second term in Eq. (\ref{centralresult}) involves the Hall conductivity tensor
\begin{equation}
    \sigma_{\mathrm{QAH}}^{i\ell} = \frac{e^2 }{2\pi \hbar} \varepsilon^{ij \ell} \hat{\bm{e}}_{j} \cdot \bm{C}_{\mathcal{V}},
    \label{QAHcond}
\end{equation}
where $\hat{\bm{e}}_{j}$ is the $j\text{th}$ Cartesian unit vector and the \textit{Chern vector} for the valence bundle $\mathcal{V}$ in three dimensions is
\begin{equation}
    \bm{C}_{\mathcal{V}} \equiv \sum_{\alpha = 1}^3 C_{\mathcal{V}}^{\alpha}\, \bm{g}_{\alpha},
\end{equation}
involving the reciprocal lattice vectors $(\bm{g}_1,\bm{g}_2,\bm{g}_3)$ defined in terms of the (possibly nonorthogonal) primitive lattice vectors $(\bm{a}_1,\bm{a}_2,\bm{a}_3)$ for the lattice $\Gamma$ through the duality relations $\bm{a}_{\alpha} \cdot\bm{g}_{\beta} = 2\pi \delta_{\alpha\beta}$. The $\alpha\text{th}$ Chern number $C_{\mathcal{V}}^{\alpha}$ is obtained by integrating the Chern class of the valence bundle over the $2$-cycle in the Brillouin zone that is orthogonal to the reciprocal lattice vector $\bm{g}_{\alpha}$. Our expression for the Hall conductivity tensor (\ref{QAHcond}) is very general, being valid for any Chern insulator in three dimensions with an arbitrary lattice structure, for example, the magnetic topological insulator $\mathrm{Mn} \mathrm{Bi}_2 \mathrm{Te}_4$ \cite{QAHExperiment6}. In the special case of a three-dimensional Chern insulator formed by stacking thin films of magnetically doped $\mathrm{Bi}_2 \mathrm{Te}_3$ or $\mathrm{Sb}_2 \mathrm{Te}_3$ along the $z$-direction \cite{QAHExperiment5}, for example, the Hall conductivity tensor in the orthogonal plane is given by
\begin{equation}
    \sigma_{\mathrm{QAH}}^{yx} = \frac{e^2}{2\pi \hbar a} C_{\mathcal{V}}^z,
    \label{sigmaHallplane}
\end{equation}
where $a$ is the spacing between planes and $C_{\mathcal{V}}^z$ is the Chern number for the $2$-cycle associated with these thin films that are extended in the $xy$-plane.

Importantly, one can show that Eq. (\ref{centralresult}) respects energy conservation. In particular, since the electric field is spatially uniform, the change in the electromagnetic energy $\mathcal{E}$ during a time interval of length $T$ is \cite{Jackson}
\begin{equation}
    \Delta \mathcal{E} = - \int_{0}^T \mathrm{d}t\, \bm{E}(t) \cdot \bm{J}(t).
\end{equation}
The zeroth-order current density vanishes, so only the first-order modification (\ref{centralresult}) contributes to $\bm{J}(t)$. Moreover, since the second term involving the quantum anomalous Hall conductivity tensor is always perpendicular to the electric field, its contribution to the quantity $\Delta \mathcal{E}$ vanishes. Also, the quantum anomalous Hall term in Eq. (\ref{centralresult}) vanishes for trivial insulators, where $C_{\mathcal{V}}^{\alpha} = 0$ for all $1 \leq \alpha \leq 3$, since the integral of the Berry curvature vanishes for these time-reversal invariant systems \cite{VanderbiltBook}.

To emphasize the causality of the first-order current density in Eq. (\ref{J1identity}), it is useful to write this expression in a different form; following some straightforward algebraic manipulations, is given by \cite{Mahon2019} 
\begin{widetext}
\begin{align}
    J^{(1)i}(\omega) = &-i \omega e^2 E^{\ell}(\omega) \sum_{mn} f_n \fint_{\mathrm{BZ}} \frac{\mathrm{d}\bm{k}}{(2\pi)^3}\, \frac{E_{m\bm{k}} - E_{n\bm{k}}}{(E_{m\bm{k}} - E_{n\bm{k}})^2 - (\hbar(\omega + i0^+))^2} \Big(\xi_{nm}^i(\bm{k}) \xi_{mn}^{\ell}(\bm{k}) + \xi_{nm}^{\ell}(\bm{k}) \xi_{mn}^{i}(\bm{k})\Big)\nonumber \\
    &- \frac{i e^2}{\hbar} E^{\ell}(\omega) \sum_{mn} f_n \fint_{\mathrm{BZ}} \frac{\mathrm{d}\bm{k}}{(2\pi)^3}\, \frac{(E_{m\bm{k}} - E_{n\bm{k}})^2}{(E_{m\bm{k}} - E_{n\bm{k}})^2 + (\hbar(\omega + i0^+))^2} \Big(\xi_{nm}^i(\bm{k}) \xi_{mn}^{\ell}(\bm{k}) - \xi_{nm}^{\ell}(\bm{k}) \xi_{mn}^i(\bm{k})\Big),
    \label{totalJalternate}
\end{align}
\end{widetext}
which can also be obtained from the corresponding expression for a metal \cite{Mahon2021} by taking the limit $f_{n\bm{k}} \to f_n$ that characterizes an insulating material; in a metal, the presence of Fermi surface means that the Fermi factor $f_{n\bm{k}}$ depends on $\bm{k} \in \mathrm{BZ}$. It can be seen from either of the expressions (\ref{J1identity}) or (\ref{totalJalternate}) that the zero-frequency limit of the full conductivity tensor is
\begin{equation}
    \lim_{\omega \to 0^+} \sigma^{i\ell}(\omega) = \sigma_{\mathrm{QAH}}^{i\ell},
\end{equation}
which is the quantum anomalous Hall effect in the presence of an applied electric field in this ``DC limit." From the expression (\ref{totalJalternate}) it is straightforward to show that the Kramers-Kronig relations are satisfied, so our first-order induced current respects causality, as expected.

\section{Discussion}\label{Sec5}
As mentioned in the Introduction, the quantum anomalous Hall response has been discussed in the ``modern theories" for finite-sized systems in the presence of static and spatially uniform electric and magnetic fields. Recall from elementary classical electromagnetism that a macroscopic polarization $\bm{P}$ in a finite sample of a material leads to a surface charge of the form $\sigma_{\text{surf}} = \bm{P}\cdot\hat{\bm{n}}$, where $\hat{\bm{n}}$ is the unit normal to the surface \cite{Jackson}. Also, there may be a surface current $\bm{K}_{\text{surf}} = c\bm{M} \times \hat{\bm{n}}$ resulting from the presence of a nonzero macroscopic orbital magnetization $\bm{M}$ in the bulk. In a two-dimensional system, the latter contributes to edge currents that circulate either clockwise or counterclockwise along the boundary, depending on the orientation of $\bm{M}$. These edge currents correspond to surface bands that cross the Fermi energy, either starting and ending in a valence band or crossing from a valence to a conduction band. Of course, in an insulating material for which time-reversal symmetry holds, only the first case is possible, since there must a ``left-right symmetry" in the surface band structure \footnote{See, \textit{e.g.}, the discussion in Ref. \cite{VanderbiltBook}}. However, in a system for which time-reversal symmetry is broken, it is possible to find band crossings of the second type too. If there is an excess number $\Delta n$ of ``up-crossings" compared to ``down-crossings," for example, then there will be a net current circulating around the boundary of the system in a fixed direction, and therefore the corresponding edge states are said to be ``chiral." It is argued in the ``modern theories" \cite{Vanderbilt2009} that this integer $\Delta n$ is actually a bulk property of the material, leading to a primitive version of the ubiquitous ``bulk-boundary correspondence" \cite{Bulkboundary1}. Thus, as the argument goes, a two-dimensional insulator with spontaneously broken time-reversal symmetry is characterized by a bulk integer that controls the number of chiral edge states on the surface. Defining a conductivity tensor through the relation $\sigma^{i\ell} = d K^i / d E_{\ell}$ in the presence of a static electric field $\bm{E}$, one obtains from the surface current $\bm{K}_{\text{surf}}$ associated with these chiral edge states a Hall conductivity tensor that --- under the assumptions leading to Eq. (\ref{sigmaHallplane}) --- is equal to ours. Since the Kubo part of our response tensor vanishes in the zero frequency limit considered by the ``modern theories," in that limit our response tensor $\sigma^{i\ell}(\omega \rightarrow 0^+)$ agrees with that of the ``modern theories."

But the arguments of the modern theories rely heavily on the presence of chiral edge states at the surface of a material, and it is unclear how to extend these arguments beyond static and spatially uniform electric fields. Moreover, they implicitly rely on the assumption that the background magnetic field is both static \textit{and} spatially uniform, since any spatial variation of the magnetic field may redistribute the charge-current distribution in such a way that it would be difficult to separate the quantum anomalous Hall conductivity from other contributions to the current density \cite{VanderbiltBook}. Of course, any system of interest in experiments is indeed of finite size and these arguments are certainly valid therein, but it is unclear how to extend these arguments identifying $\Delta n$ with a bulk quantity in the presence of optical fields, or in the presence of a spatially varying background magnetic field. An advantage of our microscopic derivation of the quantum anomalous Hall conductivity (\ref{QAHcond}) is that it \textit{is} valid for applied electric fields that are static or frequency-dependent, and for non-uniform background magnetic fields, provided they are periodic over the crystal lattice (\textit{cf.} the static vector potential in Eq. (\ref{unperturbedH})). Moreover, by implementing this microscopic derivation in a bulk crystal without resorting to chiral edge states at a surface, we circumvent the difficulties associated with surface reconstruction effects and the issue of defining a notion of ``Brillouin zone" for a finite system, as the usual construction of the Brillouin zone relies on the discrete translational symmetry of an \textit{infinite} lattice \cite{Ashcroft}.

A central feature of our formalism is the ``gauge freedom" associated with the choice of Wannier frame used in the definition of the exponentially localized Wannier functions in Eq. (\ref{unperturbedELWF}). Wannier frames are global sections of the frame bundle associated with the Bloch bundle, the existence of which follows from the trivializability of the Bloch bundle (discussed in Appendix \ref{AppendixA}). Each possible Wannier frame used in the construction of ELWFs constitutes a choice of global ``smooth gauge" for the Bloch bundle, any pair of which are related by ``gauge transformations" of the form (\ref{unitarytransform1}). The notion of a ``smooth gauge" is best understood in the framework of mathematical gauge theory \cite{HamiltonBook}, in which the frame bundle is a principal $\mathrm{U}(\mathcal{B})$-bundle over the Brillouin zone and the Bloch bundle $\mathcal{B}$ is an associated bundle thereof (through the fundamental representation). Then local (global) gauges correspond to local (global) sections of the frame bundle, and the gauge transformations are implemented by the natural right $\mathrm{U}(\mathcal{B})$-action on its total space. In this sense, the ``gauge freedom" in our formalism is naturally encoded in the principal fibre bundles associated to the Bloch bundle and its valence and conduction subbundles. 

This geometric perspective allows us to elucidate several formulae used extensively for calculations in the Appendices. For example, the global component representation (\ref{BerryconnectionWannier}) of the non-Abelian Berry connection in a Wannier frame satisfies the identity
\begin{equation}
    \varepsilon^{ijk}\partial_j \tilde{\xi}_{\alpha\beta}^{k}(\bm{k}) = i \varepsilon^{ijk} \sum_{\ell} \tilde{\xi}_{\alpha\gamma}^j(\bm{k}) \tilde{\xi}_{\gamma\beta}^{k}(\bm{k}),
\end{equation}
which is equivalent to the statement that the curvature of the Berry connection on the Bloch bundle vanishes. A similar identity holds for the local component representation (\ref{BerryconnectionBloch}) of this connection in the Bloch frame, albeit one that is only defined over open sets in the Brillouin zone. Moreover, in the derivation of Eq. (\ref{centralresult}), the details of which can be found in Appendix \ref{AppendixD}, we make use of the following local identity
\begin{equation}
    \varepsilon^{ijk}\partial_j \mathcal{W}_{mn}^k(\bm{k}) = - i\varepsilon^{ijk} \sum_{\ell} \mathcal{W}_{m\ell}^j(\bm{k}) \mathcal{W}_{\ell n}^k(\bm{k}),
    \label{Widentity}
\end{equation}
which holds for all $\bm{k} \in \mathrm{BZ} \setminus \mathscr{D}_{\mathcal{B}}$. In our geometric language, this identity is simply the Bloch-frame representation of the Cartan structure equation for the Maurer-Cartan form on $\mathrm{U}(\mathcal{B})$. The presence of such refined geometric notions is characteristic of the ``gauge freedom" in our formalism, as discussed in detail in Appendix \ref{AppendixA}.

This ``gauge freedom" manifests in a dependence of the microscopic polarization and magnetization fields (and the free charge and current densities) on our choice of Wannier frame for the Bloch bundle. But these microscopic fields featured on the right-hand-side of Eq. (\ref{microscopicpolmag}) are themselves not uniquely defined; indeed, given a pair of vector fields $\bm{a}(\bm{x},t)$ and $\bm{c}(\bm{x},t)$ and a scalar field $b(\bm{x},t)$, we can always define a new set of microscopic fields
\begin{align}
    \bm{p}'(\bm{x},t) &= \bm{p}(\bm{x},t) + \bm{\nabla}\times \bm{a}(\bm{x},t) + \bm{c}(\bm{x},t),\nonumber \\
    \bm{m}'(\bm{x},t) &= \bm{m}(\bm{x},t) - \frac{1}{c} \frac{\partial \bm{a}(\bm{x},t)}{\partial t} + \bm{\nabla}b(\bm{x},t),\nonumber \\
    \rho_F'(\bm{x},t) &= \rho_F(\bm{x},t) + \bm{\nabla}\cdot\bm{c}(\bm{x},t),\nonumber \\
    \bm{j}_F'(\bm{x},t) &= \bm{j}_F(\bm{x},t) - \frac{\partial \bm{c}(\bm{x},t)}{\partial t},
\end{align}
from which we would obtain the same electronic charge and current densities in Eq. (\ref{microscopicpolmag}). An analogous indeterminacy holds for the macroscopic expressions (\ref{macroscopicpolmag}), but in both cases the ``gauge-dependent" polarization field, magnetization field, and free charge and current densities together yield ``gauge-invariant" electronic charge and current densities. It is not then surprising that there is a ``gauge redundancy" in our microscopic treatment, since this manifests even in this simple classical analysis above. In fact, we would argue that this ``gauge freedom" is an advantage, for it can be used to minimize the spread of the ELWFs --- for example, using the Marzari-Vanderbilt spread functional \cite{Marzari2012} --- thereby improving the accuracy of approximations associated with the construction of tight-binding models \cite{CImanuscript1}, for example. 

In summary, we have implemented a formalism based on microscopic polarization and magnetization fields in extended media to derive the first-order response of a bulk Chern insulator to applied electric fields of arbitrary frequency, within the long wavelength approximation. The induced current density involves a conductivity tensor that is a sum of two terms, the first of which is a finite-frequency term that may be obtained by a Kubo analysis \cite{Kubo1}, while the second is a term associated with the quantum anomalous Hall effect and is unique to Chern insulators. This quantum anomalous Hall conductivity tensor is valid in the presence of both long-wavelength finite-frequency electric fields and time-independent background magnetic fields, the latter of which can vary in space provided they are periodic over the crystal lattice. Our work here lays the foundation for a wider research program involving the optical response of Chern insulators to generic applied electromagnetic fields in both the linear and nonlinear regimes, and the extension of this approach to treat the optical response at surfaces.

\begin{acknowledgements}
    We thank Perry Mahon for useful discussions. This work was supported by the Natural Sciences and Engineering Research Council of Canada (NSERC). J. G. K. acknowledges an Ontario Graduate Scholarship. 
\end{acknowledgements}

\appendix

\section{Mathematical background}\label{AppendixA}

Here we summarize the fundamental geometric structures and topological invariants that underlie our discussion of Chern insulators in Sec. \ref{Sec2}.

\subsection{Band structure and Hilbert bundles}

The systems we consider here are $d$-dimensional \textit{bulk crystals}, by which we mean a countable subset $\mathscr{C}$ of the affine space $\mathbb{A}^d$ (modelled on the vector space $\mathbb{R}^d$) that is equipped with an action of a lattice $\Gamma$ of discrete translations \cite{Freed}. The points in $\mathscr{C}$ are the locations of ion cores in the crystal, and the action of $\Gamma$ on $\mathscr{C}$ defines an equivalence relation thereon, where any pair of points in $\mathscr{C}$ are elements of the same equivalence class in $\mathscr{C}/\Gamma$ if and only if they differ by addition of a lattice site $\bm{R} \in \Gamma$. The cardinality of the quotient set $\mathscr{C}/\Gamma$ is the number of distinct sublattices in $\Gamma$, and is equal to unity if $\Gamma$ is a Bravais lattice \cite{Fruchart1}. Defining the reciprocal lattice $\Gamma^*$ in the usual way, the quotient of the space of wavevectors $\bm{k} \in \mathbb{R}^d$ by the additive action of $\Gamma^*$ thereon is equipped with the quotient topology, and the result is topologically equivalent to a $d$-torus \cite{Carpentier2017}. There is a unique smooth structure on this quotient space obtained from a discrete version of the ``quotient manifold theorem" \cite{LeeBook}, turning it into a smooth manifold,
the \textit{Brillouin zone} $\mathrm{BZ}$. 

For each point $\bm{k} \in\mathrm{BZ}$, the collection of cell-periodic Bloch states $\{\ket{n\bm{k}}\}_{n\in \mathbb{N}}$ span a separable Hilbert space 
\begin{equation}
    \mathcal{B}_{\bm{k}} \equiv \text{span}_{\mathbb{C}}\big(\{\ket{n\bm{k}}\}_{n\in \mathbb{N}}\big)
    \label{hilb1}
\end{equation}
equipped with the Hermitian inner product
\begin{equation}
    \big(n\bm{k}\big|m\bm{k}\big) = \frac{1}{\Omega_{\text{uc}}} \int_{\Omega} \mathrm{d}\bm{x}\, u_{n\bm{k}}^*(\bm{x}) u_{m\bm{k}}(\bm{x}),
    \label{innerprod1}
\end{equation}
where $\Omega$ is the Wigner-Seitz cell and $\Omega_{\text{uc}}$ is its volume. Since the Hilbert spaces $\mathcal{B}_{\bm{k}}$ have the same dimension for all $\bm{k} \in \mathrm{BZ}$, we can define a vector bundle $\mathcal{B} \overset{\pi_{\mathcal{B}}}{\longrightarrow} \mathrm{BZ}$ whose fibre above a point $\bm{k} \in \mathrm{BZ}$ is the complex vector space $\mathcal{B}_{\bm{k}}$. The inner product (\ref{innerprod1}) on the fibres $\mathcal{B}_{\bm{k}}$ can be smoothly extended to a Hermitian metric $h$ on this complex vector bundle and so the pair $(\mathcal{B},h)$ is a smooth Hilbert bundle over the manifold $\mathrm{BZ}$ called the \textit{Bloch bundle} \cite{Panati2017}; to simplify notation, we denote this Hilbert bundle by its total space $\mathcal{B}$. 

We focus on \textit{band insulators} for which there is a well-defined bandgap across the Brillouin zone separating the first $N \in \mathbb{N}$ occupied ``valence" bands from the remaining unoccupied ``conduction" bands. Associated with the Hamiltonian (\ref{unperturbedH}) are a family of single-particle Hamiltonians $\hat{H}_0(\bm{k})$ parametrized by $\bm{k} \in \mathrm{BZ}$ and given by 
\begin{equation}
    \hat{H}_0(\bm{k}) = \frac{1}{2m}\Big(\hat{\bm{p}} - \frac{e}{c} \bm{a}_{\text{static}}(\hat{\bm{x}}) + \hbar \bm{k}\Big)^2 + \mathrm{V}_{\Gamma}(\hat{\bm{x}}),
\end{equation}
where $\hat{\bm{x}}$ and $\hat{\bm{p}}$ are the first-quantized position and momentum operators. This guarantees that we have
\begin{equation}
    \hat{H}_0(\bm{k})\ket{n\bm{k}} = E_{n\bm{k}}\ket{n\bm{k}}, 
\end{equation}
and that the map $\bm{k} \mapsto \hat{H}_0(\bm{k})$ is globally smooth. The existence of the bandgap means that we can define an orthogonal projector at each $\bm{k} \in \mathrm{BZ}$ through the Riesz formula \cite{Brouder}
\begin{equation}
    \hat{P}_{\bm{k}} = \frac{1}{2\pi i} \int_{\mathscr{C}_{\bm{k}}} \mathrm{d}z\, \Big(\hat{H}_0(\bm{k}) - z \mathbb{I}\Big)^{-1}
    \label{Riesz}
\end{equation}
in such a way that the map $\bm{k} \mapsto \hat{P}_{\bm{k}}$ is globally smooth. Here $\mathscr{C}_{\bm{k}}$ is a positively-oriented contour in $\mathbb{C}$ enclosing the eigenvalues $\{E_{n\bm{k}}\}_{n=1}^N$ and no others. At each point $\bm{k} \in \mathrm{BZ}$ this projector maps elements of the fibre $\mathcal{B}_{\bm{k}}$ onto the subspace $\mathcal{V}_{\bm{k}}$ spanned by the cell-periodic Bloch states $\ket{n\bm{k}}$ for all $1 \leq n \leq N$. Since this is a finite-dimensional subspace of $\mathcal{B}_{\bm{k}}$, it admits a well-defined orthogonal complement that we denote by $\mathcal{C}_{\bm{k}}$. The smooth map $\bm{k} \mapsto \hat{P}_{\bm{k}}$ is of fixed rank across the Brillouin zone and therefore induces a Whitney-sum decomposition \cite{NakaharaBook}
\begin{equation}
    \mathcal{B} = \mathcal{V} \oplus \mathcal{C} \overset{\pi_{\mathcal{B}}}{\longrightarrow} \mathrm{BZ},
\end{equation}
where $\mathcal{V} \overset{\pi_{\mathcal{V}}}{\longrightarrow} \mathrm{BZ}$ is called the \textit{valence bundle} and encodes the spectral properties of the occupied bands, while $\mathcal{C} \overset{\pi_{\mathcal{C}}}{\longrightarrow} \mathrm{BZ}$ is called the \textit{conduction bundle} and encodes the spectral properties of the unoccupied bands. The fibres of the conduction bundle are the images under a projector of the form (\ref{Riesz}), but where the contour encloses the eigenvalues of the conduction bands at $\bm{k}$. Both the valence and conduction bundles are smooth Hilbert bundles \cite{Fruchart1}; they are examples of \textit{projected subbundles} of the Bloch bundle, since they are defined through a smooth family of projectors (\ref{Riesz}) on the fibres thereof.

\subsection{Smooth gauges and frame bundles}
The Bloch bundle and its valence and conduction subbundles encode the spectral properties of the band structure of a crystalline insulator. The Bloch bundle is a Hilbert bundle of infinite rank and is therefore \textit{trivializable}, meaning that it is isomorphic to the trivial bundle $\mathrm{BZ} \times \mathcal{B}_0$ for some fixed Hilbert space $\mathcal{B}_0$ \cite{Freed}. Such a choice of isomorphism is called a \textit{trivialization} for $\mathcal{B}$, which is equivalent to a choice of smooth map $\bm{k} \mapsto (\ket{\alpha\bm{k}})_{\alpha\in\mathbb{N}}$ assigning to every $\bm{k} \in \mathrm{BZ}$ an orthonormal Schauder basis $(\ket{\alpha\bm{k}})_{\alpha \in \mathbb{N}}$ for the fibre $\mathcal{B}_{\bm{k}}$ of the Bloch bundle. This smooth map is called a \textit{global frame} for $\mathcal{B}$; in Sec. \ref{Sec2} we referred to such a (smooth) global frame as a ``Wannier frame,"  since the states therein are used in the construction of ELWFs. There are many possible choices for such a Wannier frame, and this ``gauge freedom" is encoded in the so-called ``frame bundle" associated to the Bloch bundle.

We denote by $F\mathcal{B}_{\bm{k}}$ the set of orthonormal Schauder bases for the fibre $\mathcal{B}_{\bm{k}}$ at each $\bm{k}\in\mathrm{BZ}$. Examples include the basis of cell-periodic Bloch states $(\ket{n\bm{k}})_{n\in \mathbb{N}}$ and any of the Wannier bases $(\ket{\alpha\bm{k}})_{\alpha \in \mathbb{N}}$ described in Sec. \ref{Sec2}. By fixing a particular basis, one can show that elements of $F\mathcal{B}_{\bm{k}}$ are in one-to-one correspondence with elements of the Lie group $\mathrm{U}(\mathcal{B}_{\bm{k}})$ of unitary operators on $\mathcal{B}_{\bm{k}}$ \cite{Freed}. The topology and smooth structure on $\mathrm{U}(\mathcal{B}_{\bm{k}})$ are then inherited by the set $F\mathcal{B}_{\bm{k}}$ under this correspondence, turning it into a smooth manifold at each $\bm{k} \in \mathrm{BZ}$ \cite{Kriegl}. Thus, we can construct a smooth fibre bundle over the Brillouin zone whose total space is the disjoint union
\begin{equation}
    F\mathcal{B} \equiv \coprod_{\bm{k} \in \text{BZ}} F\mathcal{B}_{\bm{k}},
\end{equation}
called the \textit{frame bundle} for $\mathcal{B}$. Through standard constructions \cite{Kriegl} this frame bundle becomes a principal fibre bundle with the structure group $\mathrm{U}(\mathcal{B})$, which is the Lie group of unitary transformations on the fibres of the Bloch bundle. Sections of the frame bundle $F\mathcal{B}$ are smooth frames over open sets in the Brillouin zone, examples of which include the locally-defined Bloch frame $\bm{k}\mapsto (\ket{n\bm{k}})_{n\in\mathbb{N}}$ and any of the globally-defined Wannier frames $\bm{k}\mapsto(\ket{\alpha\bm{k}})_{\alpha \in \mathbb{N}}$. The frame bundle $F\mathcal{B}$ comes equipped with a right $\mathrm{U}(\mathcal{B})$-action that implements the local unitary transformation (\ref{unitarytransform1}) between elements of these Bloch and Wannier frames, as well as between different Wannier frames.

We can similarly define the frame bundles $F\mathcal{V}$ and $F\mathcal{C}$ associated to the valence and conduction bundles, both of which are principal fibre bundles over the Brillouin zone with the structure groups $\mathrm{U}(\mathcal{V})$ and $\mathrm{U}(\mathcal{C})$, respectively. Since the Bloch bundle is trivializable, its frame bundle is also trivializable as a principal $\mathrm{U}(\mathcal{B})$-bundle \cite{Freed}, where in this case a trivialization is an isomorphism between $F\mathcal{B}$ and the trivial bundle $\text{BZ} \times \mathrm{U}(\mathcal{B})$. Such an isomorphism can be implemented by a choice of globally-defined smooth section of $F\mathcal{B}$, that is, a global frame for the Bloch bundle, often referred to in band theory as a choice of (global) \textit{smooth gauge} \cite{VanderbiltBook}. But the valence and conduction bundles of a Chern insulator are \textit{not} trivializable, and so there do not exist smooth global frames (or smooth gauges) for these Hilbert bundles. However, since they are locally trivial it follows that there always exist \textit{local} smooth gauges for these Hilbert bundles, that is, local smooth sections of their respective frame bundles.

\subsection{The non-Abelian Berry connection}
Being a Hilbert bundle that is trivializable, there is a natural choice of connection on the Bloch bundle \cite{Freed}, which in turn induces connections on its valence and conduction subbundles. To derive these connections, we note that (being projected subbundles of the Bloch bundle) they are defined through smooth projection maps of the form $\bm{k} \mapsto \hat{P}_{\bm{k}}$. Taking the exterior derivative of the projection identity $\hat{P}_{\bm{k}}^2 = \hat{P}_{\bm{k}}$, one can show that
\begin{equation}
    \mathrm{d}\hat{P}_{\bm{k}} + [\hat{A}_{\bm{k}}, \hat{P}_{\bm{k}}] = 0,
    \label{projectorcondition1}
\end{equation}
where we have defined the $1$-form 
\begin{equation}
    \hat{A}_{\bm{k}} \equiv - [\mathrm{d}\hat{P}_{\bm{k}}, \hat{P}_{\bm{k}}].\label{connect1form}
\end{equation}
This is a skew--self adjoint linear operator on the fibre $\mathcal{B}_{\bm{k}}$ of the Bloch bundle and is therefore an element of the Lie algebra $\mathfrak{u}(\mathcal{B})$ of $\mathrm{U}(\mathcal{B})$. Consider the endomorphism bundle $\mathrm{End}(\mathcal{B})$ of the Bloch bundle $\mathcal{B}$ \cite{LeeBook}, the sections of which are smooth maps $\bm{k}\mapsto \hat{X}_{\bm{k}}$ of self-adjoint (continuous) linear operators $\hat{X}_{\bm{k}}$ on the fibres $\mathcal{B}_{\bm{k}}$ of $\mathcal{B}$. Defining a connection $\nabla^{\mathcal{B}}$ on $\mathcal{B}$ is equivalent to defining a connection $\hat{\nabla}^{\mathcal{B}}$ on its endomorphism bundle \cite{Baez}, which acts on sections thereof by
\begin{equation}
    \hat{\nabla}^{\mathcal{B}} \hat{X}_{\bm{k}} = \mathrm{d}\hat{X}_{\bm{k}} + \big[\hat{A}_{\bm{k}}, \hat{X}_{\bm{k}}\big],
\end{equation}
where $\hat{A}_{\bm{k}}$ is the connection $1$-form. And so we implement the projection identity (\ref{projectorcondition1}) by taking this connection $1$-form to be precisely the $\mathfrak{u}(\mathcal{B})$-valued $1$-form in Eq. (\ref{connect1form}), meaning that $\hat{\nabla}^{\mathcal{B}} \hat{P}_{\bm{k}} = 0$ for all $\bm{k} \in \mathrm{BZ}$. Then the associated connection on the Bloch bundle acts on smooth sections $\bm{k}\mapsto \Psi_{\bm{k}}$ thereof by \cite{Baez}
\begin{equation}
    \nabla^{\mathcal{B}}\Psi_{\bm{k}} = \mathrm{d}\Psi_{\bm{k}} + \hat{A}_{\bm{k}}\Psi_{\bm{k}},
    \label{abstractBerryconnection}
\end{equation}
from which follows
\begin{equation}
    \hat{Q}_{\bm{k}} \nabla^{\mathcal{B}}\Psi_{\bm{k}}^{\mathcal{V}} = 0
\end{equation}
for any section $\bm{k}\mapsto \Psi_{\bm{k}}^{\mathcal{V}}$ of the valence bundle $\mathcal{V}$. This means that the connection $\nabla^{\mathcal{B}}$ takes sections of $\mathcal{V}$ to $\mathcal{V}$-valued $1$-forms, as is required for a connection on $\mathcal{V}$ \cite{Baez}. And, indeed, defining $\nabla^{\mathcal{V}} \equiv \hat{P}\, \nabla^{\mathcal{B}}$ we find
\begin{equation}
    \nabla^{\mathcal{V}} \Psi_{\bm{k}}^{\mathcal{V}} = \hat{P}_{\bm{k}} \mathrm{d} \Psi_{\bm{k}}^{\mathcal{V}}
\end{equation}
for any section of $\mathcal{V}$, often written as $\nabla^{\mathcal{V}} = \hat{P}\mathrm{d}$ \cite{Brouder}. 

The above holds for any projected subbundle of the Bloch bundle, but by setting $\hat{P}_{\bm{k}} = \mathbb{I}$ we can address the entire Bloch bundle as well, over which we thereby obtain the trivial connection $\nabla^{\mathcal{B}} = \mathrm{d}$ on the Bloch bundle $\mathcal{B}$. This is called the \textit{non-Abelian Berry connection} on the Bloch bundle, and the projected connection $\nabla^{\mathcal{V}}$ is the associated Berry connection on the valence bundle \cite{Fruchart1}. 

Given a frame for the Bloch bundle, one can define an associated ``component representation" of the connection (\ref{abstractBerryconnection}) in that frame. For example, choosing a Wannier frame $(\ket{\alpha\bm{k}})_{\alpha \in \mathbb{N}}$ for the Bloch bundle, the global component representation of this connection is
\begin{equation}
    \tilde{\xi}_{\alpha\beta}^{\nabla}(\bm{k}) = i \big(\alpha\bm{k}|\nabla^{\mathcal{B}}\beta\bm{k}\big),
\end{equation}
where the factor of $i$ is included so that this $1$-form is valued in the endomorphism bundle \cite{Brouder, Fruchart1}. If we instead choose the local cell-periodic Bloch frame $(\ket{n\bm{k}})_{n\in\mathbb{N}}$, then the \textit{local} component representation of this connection is
\begin{equation}
    \xi_{mn}^{\nabla}(\bm{k}) = i\big(m\bm{k}|\nabla^{\mathcal{B}}n\bm{k}\big).
\end{equation}
In particular, if we take $\hat{P}_{\bm{k}} = \mathbb{I}$ so that we are working with the trivial connection $\nabla^{\mathcal{B}} = \mathrm{d}$ on the Bloch bundle, then the global component representation of this Berry connection in a Wannier frame is
\begin{equation}
    \tilde{\xi}_{\alpha\beta}(\bm{k}) = \tilde{\xi}_{\alpha\beta}^i(\bm{k})\mathrm{d}k^i,
\end{equation}
where the component functions are defined in Eq. (\ref{BerryconnectionWannier}), and the local component representation of this Berry connection in the cell-periodic Bloch frame is
\begin{equation}
    \xi_{mn}(\bm{k}) = \xi_{mn}^i(\bm{k}) \mathrm{d}k^i,
\end{equation}
where the component functions are defined in Eq. (\ref{BerryconnectionBloch}).

The component functions featured in these two representations of the non-Abelian Berry connection are related locally by Eq. (\ref{BlochWannierconnection}). Up to a factor of $i$, the second term in Eq. (\ref{BlochWannierconnection}) involves the matrix components of the \textit{Maurer-Cartan form} on $\mathrm{U}(\mathcal{B})$ \cite{NakaharaBook, Panati2017}, given by
\begin{equation}
    \mathcal{W}_{\bm{k}} = \big[i\big(\partial_i U_{\bm{k}}\big)U_{\bm{k}}^{-1}\big] \mathrm{d}k^i,
    \label{Winverse}
\end{equation}
where the quantity in parentheses is the Hermitian matrix $\mathcal{W}^a(\bm{k})$. The Maurer-Cartan form on $\mathrm{U}(\mathcal{B})$ satisfies the \textit{Cartan structure equation} \cite{NakaharaBook}, which in this case is
\begin{equation}
    \mathrm{d}\mathcal{W}_{\bm{k}} + i \mathcal{W}_{\bm{k}}\wedge_{\mathcal{B}} \mathcal{W}_{\bm{k}} = 0,
\end{equation}
whose local component representation in the Bloch frame is given by Eq. (\ref{Widentity}). 

\section{Integration over the Brillouin zone}\label{AppendixA4}
The ``adapted integrals" over the manifold $\mathrm{BZ}$ of the form introduced in Eq. (\ref{Chernnumber2d}) are necessary to be precise about two issues in Brillouin-zone integration. First, it is often the case that the integrand is only defined locally on open subsets of the manifold $\mathrm{BZ}$, such as in the Bloch-frame representations of the non-Abelian Berry connection (\ref{BerryconnectionBloch}) and the Maurer-Cartan form (\ref{Wmatrix}), even away from band crossings. This can be resolved using standard techniques from differential geometry \cite{LeeBook}: Consider an open cover $\{U_{\alpha}\}_{\alpha}$ of the smooth manifold $\mathrm{BZ}$ together a partition of unity $\{p_{\alpha}\}_{\alpha}$ subordinate to this cover. The integral of some function $f$ over the $d$-dimensional Brillouin zone is
\begin{equation}
    \int_{\mathrm{BZ}} f \equiv \sum_{\alpha} \int_{U_{\alpha}} f_{\alpha},
    \label{int1}
\end{equation}
where we have defined $f_{\alpha} : U_{\alpha} \to \mathbb{R}$ by $f_{\alpha} \equiv p_{\alpha} f$, so that if $f$ is only piecewise defined over open sets in $\mathrm{BZ}$, then we can find an open cover that is sufficiently fine so that the right-hand-side above  is well-defined. 

However, there are still subtleties in the definition of such an integral owing to band crossings at points in the locus of degeneracies $\mathscr{D}_{\mathcal{B}}$ \cite{Kaufmann}. Label those points in $\mathscr{D}_{\mathcal{B}}$ that are also in the open set $U_{\alpha}$ by $\bm{k}_{i_{\alpha}} \in \mathscr{D}_{\mathcal{B}} \cap U_{\alpha}$, and enclose each of these points by a \textit{closed} ball $B_{\varepsilon}(\bm{k}_{i_{\alpha}})$ with radius $\varepsilon > 0$ chosen to be sufficiently small so that $B_{\varepsilon}(\bm{k}_{i_{\alpha}}) \subseteq U_{\alpha}$. Define the union set
\begin{equation}
    \mathcal{N}_{\alpha}(\varepsilon) \equiv \bigcup_{i_{\alpha}} B_{\varepsilon}(\bm{k}_{i_{\alpha}}),
\end{equation}
in which case using Eq. (\ref{int1}) we can write
\begin{equation}
    \int_{\mathrm{BZ}} f = \lim_{\varepsilon\to 0^+} \sum_{\alpha} \left(\int_{U_{\alpha}\setminus \mathcal{N}_{\alpha}(\varepsilon)} f_{\alpha} + \int_{\mathcal{N}_{\alpha}(\varepsilon)} f_{\alpha}\right).\label{int2}
\end{equation}
Then given the (closed) set $\mathcal{N}_{\varepsilon} \equiv \cup_{\alpha}\, \mathcal{N}_{\alpha}(\varepsilon)$, which is a tubular neighborhood that deformation retracts onto $\mathscr{D}_{\mathcal{B}}$ (in the limit $\varepsilon \to 0^+$) \cite{Kaufmann}, we define
\begin{equation}
    \fint_{\mathrm{BZ}} f \equiv \lim_{\varepsilon\to 0^+} \left(\int_{\mathrm{BZ}\setminus \mathcal{N}_{\varepsilon}} f + \int_{\mathcal{N}_{\varepsilon}}f \right),
\end{equation}
which is independent of the choice of open cover for the manifold $\mathrm{BZ}$ and partition of unity subordinate thereto, but is equivalent to Eq. (\ref{int2}) when such a choice is made. By defining these ``adapted integrals" we circumvent problems associated with integration across the $\mathrm{BZ}$ in the presence of points and lines of degeneracy; planes of degeneracy that split the Brillouin are more complicated and are not considered here. Evaluation of the second integral in the expression above must be done on a case-by-case basis, and these ``adapted integrals" can also be defined for differential forms over the $\mathrm{BZ}$ and even over embedded submanifolds thereof.

\section{First-order modification of the SPDM}\label{AppendixB}
The full SPDM satisfies the differential equation \cite{Mahon2019}
\begin{equation}
    i \hbar \frac{\partial \eta_{\alpha\bm{R}'';\beta\bm{R}'}(t)}{\partial t} = \sum_{\mu\nu\bm{R}_1\bm{R}_2} \mathfrak{F}_{\alpha\bm{R}'';\beta\bm{R}'}^{\mu\bm{R}_1;\nu\bm{R}_2}(t) \eta_{\mu\bm{R}_1;\nu\bm{R}_2}(t),
\end{equation}
where in the presence of a spatially uniform electric field, and neglecting any magnetic effects, the quantity on the right-hand-side is given by
\begin{align}
    \mathfrak{F}_{\alpha\bm{R}'';\beta\bm{R}'}^{\mu\bm{R}_1;\nu\bm{R}_2}(t) &= \delta_{\nu\beta} \delta_{\bm{R}_2\bm{R}'} H_{\alpha\bm{R}'';\mu\bm{R}_1}(\bm{R}_a,t)\nonumber \\
    &- \delta_{\mu\alpha} \delta_{\bm{R}_1\bm{R}''}H_{\nu\bm{R}_2;\beta\bm{R}'}(\bm{R}_a,t),
\end{align}
involving hopping matrix elements $H_{\alpha\bm{R}'';\beta\bm{R}'}(\bm{R}_a,t)$ of the minimal coupling Hamiltonian \cite{Mahon2020}. Here the lattice site $\bm{R}_a \in \Gamma$ is arbitrary and will be fixed later in the calculation. Expanding these quantities in powers of the applied electric field leads to
\begin{equation}
    \mathfrak{F}_{\alpha\bm{R}'';\beta\bm{R}'}^{\mu\bm{R}_1;\nu\bm{R}_2}(t) = \mathfrak{F}_{\alpha\bm{R}'';\beta\bm{R}'}^{(0)\mu\bm{R}_1;\nu\bm{R}_2} + \mathfrak{F}_{\alpha\bm{R}'';\beta\bm{R}'}^{(1)\mu\bm{R}_1;\nu\bm{R}_2}(t) + \dots,
\end{equation}
which, together with the analogous expansion (\ref{expansionSPDM}) of the SPDM, results in the first-order differential equation
\begin{align}
    i \hbar \frac{\partial \eta_{\alpha\bm{R}'';\beta\bm{R}'}^{(1)}(t)}{\partial t} &= \sum_{\mu\nu\bm{R}_1\bm{R}_2} \mathfrak{F}_{\alpha\bm{R}'';\beta\bm{R}'}^{(0)\mu\bm{R}_1;\nu\bm{R}_2} \eta_{\mu\bm{R}_1;\nu\bm{R}_2}^{(1)}(t)\nonumber \\
    &+ \sum_{\mu\nu\bm{R}_1\bm{R}_2} \mathfrak{F}_{\alpha\bm{R}'';\beta\bm{R}'}^{(1)\mu\bm{R}_1;\nu\bm{R}_2}(t) \eta_{\mu\bm{R}_1;\nu\bm{R}_2}^{(0)},
\end{align}
where the zeroth-order SPDM is given in Eq. (\ref{unperturbedSPDM}). Introducing a Fourier decomposition (\ref{Fourieranalysis}) of both sides and using the Bloch-frame identity
\begin{equation}
    \eta_{m\bm{k};n\bm{k}'}(\omega) \equiv \sum_{\alpha\beta\bm{R}'\bm{R}''} \bra{\psi_{m\bm{k}}}\ket{\alpha\bm{R}''} \eta_{\alpha\bm{R}'';\beta\bm{R}'}(\omega) \bra{\beta\bm{R}'}\ket{\psi_{n\bm{k}'}},
    \label{BlochframeSPDM1}
\end{equation}
the first-order SPDM is given by
\begin{widetext}
\begin{align}
    \eta_{m\bm{k};n\bm{k}'}^{(1)}(\omega) = - \sum_{\mu\nu\bm{R}_1\bm{R}_2} \sum_{\alpha\beta\bm{R}'\bm{R}''} \frac{\bra{\psi_{m\bm{k}}}\ket{\alpha\bm{R}''}\mathfrak{F}_{\alpha\bm{R}'';\beta\bm{R}'}^{(1)\mu\bm{R}_1;\nu\bm{R}_2}(\omega) \eta_{\mu\bm{R}_1;\nu\bm{R}_2}^{(0)} \bra{\beta\bm{R}'}\ket{\psi_{n\bm{k}'}}}{E_{m\bm{k}} - E_{n\bm{k}'} - \hbar (\omega + i0^+)},
    \label{firstorderSPDMBloch}
\end{align}
where the first-order quantity on the right-hand-side is
\begin{align}
    \mathfrak{F}_{\alpha\bm{R}'';\beta\bm{R}'}^{(1)\mu\bm{R}_1;\nu\bm{R}_2}(\omega) = 
    \delta_{\nu\beta} \delta_{\bm{R}_2\bm{R}'} H_{\alpha\bm{R}'';\mu\bm{R}_1}^{(1)}(\bm{R}_a,\omega) - \delta_{\mu\alpha} \delta_{\bm{R}_1\bm{R}''} H_{\nu\bm{R}_2;\beta\bm{R}'}^{(1)}(\bm{R}_a,\omega)
\end{align}
with the first-order hopping matrix elements given by
\begin{equation}
    H_{\mu\bm{R}_1;\nu\bm{R}_2}^{(1)}(\bm{R}_a,\omega) = - e E^{\ell}(\omega) \int \mathrm{d}\bm{x}\, W_{\mu\bm{R}_1}^*(\bm{x})(x^{\ell} - R_a^{\ell}) W_{\nu\bm{R}_2}(\bm{x}).
    \label{firstorderhopping}
\end{equation}
Inserting these expressions into Eq. (\ref{firstorderSPDMBloch}) and implementing the inverse transformation of Eq. (\ref{BlochframeSPDM1}), we find the first-order modification of the single-particle density matrix is given by
\begin{align}
    \eta_{\alpha\bm{R}'';\beta\bm{R}'}^{(1)}(\omega) = - \sum_{\mu\nu\bm{R}_1\bm{R}_2} \sum_{mn} f_{nm} \int_{\text{BZ}} \mathrm{d}\bm{k}\mathrm{d}\bm{k}'\, \frac{\bra{\alpha\bm{R}''}\ket{\psi_{m\bm{k}}}\bra{\psi_{m\bm{k}}}\ket{\mu\bm{R}_1}H_{\mu\bm{R}_1;\nu\bm{R}_2}^{(1)}(\bm{R}_a,\omega) \bra{\nu\bm{R}_2}\ket{\psi_{n\bm{k}'}}\bra{\psi_{n\bm{k}'}}\ket{\beta\bm{R}'}}{E_{m\bm{k}} - E_{n\bm{k}'} - \hbar(\omega + i0^+)}.
\end{align}
A tedious but straightforward calculation using the identity \cite{Marzari2012}
\begin{equation}
    \int \mathrm{d}\bm{x}\, W_{\mu\bm{R}}^*(\bm{x})x^{i} W_{\nu\bm{0}}(\bm{x}) = \frac{\Omega_{\text{uc}}}{(2\pi)^3} \int_{\text{BZ}}\mathrm{d}\bm{k}\, e^{i\bm{k}\cdot\bm{R}} \tilde{\xi}_{\mu\nu}^{i}(\bm{k}),
    \label{usefulidentity1}
\end{equation}
leads to the desired result given in Eq. (\ref{firstorderSPDMWannier}).
\end{widetext}

\section{Polarization and magnetization, and free charges and currents}\label{AppendixC}
Here we discuss the quantities and results of Sec. \ref{Sec3}. We give microscopic expressions for the site polarization and magnetization fields, and the corresponding zeroth-order expressions for the macroscopic polarization and magnetization derived previously \cite{CImanuscript1}. We then outline the derivation of the first-order contribution to the macroscopic polarization field and free current density. 

\subsection{Site quantities}
As we have motivated in previous work \cite{CImanuscript1}, the site polarization field at the lattice site $\bm{R}$ is taken to be
\begin{equation}
    \bm{p}_{\bm{R}}(\bm{x},t) = \int \mathrm{d}\bm{y}\, \bm{s}(\bm{x};\bm{y},\bm{R}) \Big(\rho_{\bm{R}}^e(\bm{y},t) + \rho_{\bm{R}}^{\text{ion}}(\bm{y})\Big),\label{micropolarization}
\end{equation}
where the ``relator" is defined in Eq. (\ref{relatorS}). Meanwhile, the site magnetization field is a sum of two terms, the first of which is an ``atomic-like" contribution
\begin{equation}
    \bar{m}_{\bm{R}}^i(\bm{x},t) = \frac{1}{c} \int \mathrm{d}\bm{y}\, \alpha^{ij}(\bm{x};\bm{y},\bm{R}) j_{\bm{R}}^j(\bm{y},t),
\end{equation}
where we have introduced another ``relator" \cite{Mahon2019}
\begin{equation}
    \alpha^{ij}(\bm{x};\bm{y},\bm{R}) = \varepsilon^{imn} \int_{C(\bm{y},\bm{R})} \mathrm{d}z^m\, \frac{\partial z^n}{\partial y^j} \delta(\bm{x}-\bm{z}),\label{relatoralpha}
\end{equation}
while the second term arises because there is a local non-conservation of the charge-current distribution at each lattice site $\bm{R}$, leading to an ``itinerant" contribution 
\begin{equation}
    \tilde{m}_{\bm{R}}^i(\bm{x},t) = \frac{1}{c} \int \mathrm{d}\bm{y}\, \alpha^{ij}(\bm{x};\bm{y},\bm{R}) \tilde{j}_{\bm{R}}^j(\bm{y},t),
\end{equation}
involving an itinerant ``site" current density $\tilde{\bm{j}}_{\bm{R}}(\bm{x},t)$ defined in previous work \cite{Mahon2019}. The total site magnetization field is then 
\begin{align}
    \bm{m}_{\bm{R}}(\bm{x},t) = \bar{\bm{m}}_{\bm{R}}(\bm{x},t) + \tilde{\bm{m}}_{\bm{R}}(\bm{x},t),
    \label{total_m}
\end{align}
plus in general a spin contribution were spin effects to be included.  We neglect those here, and so the ``orbital magnetization" (\ref{total_m}) is all that we consider. Choosing a straight-line path for the relators in Eqs. (\ref{relatorS}) and (\ref{relatoralpha}), and implementing a formal Taylor expansion of the resulting expressions, we thereby obtain microscopic electric and magnetic multipole expansions of the form
\begin{align}
    p_{\bm{R}}^i(\bm{x},t) &= \mu_{\bm{R}}^i(t) \delta(\bm{x}-\bm{R}) - q_{\bm{R}}^{ij}(t) \frac{\partial \delta(\bm{x}-\bm{R})}{\partial x^j} + \dots,\nonumber \\
    m_{\bm{R}}^i(\bm{x},t) &= \nu_{\bm{R}}^i(t) \delta(\bm{x}-\bm{R}) + \dots,\label{micromultipoleexpansions}
\end{align}
where the site electric and magnetic dipole moments are given in Eq. (\ref{dipolemoments}).

\subsection{Zeroth-order and first-order quantities}
As we have shown previously \cite{CImanuscript1}, the zeroth-order contribution to the macroscopic polarization, obtained from the zeroth-order SPDM (\ref{unperturbedSPDM}), is given by
\begin{equation}
    P^{(0)i} = e\sum_{n}f_n\fint_{\text{BZ}}\frac{\mathrm{d}\boldsymbol{k}}{(2\pi)^3}\Big(\xi^i_{nn}(\bm{k})+\mathcal{W}^i_{nn}(\bm{k})\Big) + P_{\text{ion}}^i,\label{mu}
\end{equation}
where the quantities $\xi_{mn}^i(\bm{k})$ and $\mathcal{W}_{mn}^i(\bm{k})$ are defined in Eqs. (\ref{BerryconnectionBloch}) and (\ref{Wmatrix}), respectively. This agrees with the polarization introduced in the ``modern theory." The zeroth-order contribution to the orbital magnetization is
\begin{widetext}
\begin{align}
    M^{(0)i} = \frac{e}{2\hbar c} \varepsilon^{ijk} \sum_{mn} \fint_{\text{BZ}}\frac{\mathrm{d}\boldsymbol{k}}{(2\pi)^3} \bigg(f_n E_{m\bm{k}} \Big( \delta_{mn} \partial_j \xi^k_{nn}(\bm{k}) - \text{Im}\big[\xi^j_{nm}(\bm{k})\xi^k_{mn}(\bm{k})\big]\Big) + f_{nm} E_{n\bm{k}} \, \text{Im}\big[\mathcal{W}^j_{nm}(\bm{k})\mathcal{W}^k_{mn}(\bm{k})\big]\bigg),\label{nu}
\end{align}
where the contributions from the first two terms in the integrand lead to the expression for the ground-state magnetization of a trivial insulator in the ``modern theory," while the third term is a novel ``gauge-dependent" contribution unique to Chern insulators \cite{CImanuscript1}; yet it differs from the extra contribution that has been suggested for Chern insulators within the ``modern theories" \cite{Vanderbilt2009}.  

To obtain the first-order contribution to the macroscopic polarization field, we combine Eq. (\ref{micropolarization}) for the site polarization field together with the first-order form of the microscopic charge density in Eq. (\ref{siterhoJ}), leading to
\begin{equation}
    p_{\bm{R}}^{(1)i}(\bm{x},\omega) = \sum_{\alpha\beta\bm{R}'\bm{R}''} \left[\int \mathrm{d}\bm{y}\, s^i(\bm{x};\bm{y},\bm{R}) \rho_{\beta\bm{R}';\alpha\bm{R}''}^{(0)}(\bm{y},\bm{R})\right] \eta_{\alpha\bm{R}'';\beta\bm{R}'}^{(1)}(\omega),
\end{equation}
where the zeroth-order ``generalized site-quantity matrix element" for the electronic charge density is
\begin{equation}
    \rho_{\beta\bm{R}';\alpha\bm{R}''}^{(0)}(\bm{x},\bm{R}) = \frac{e}{2} (\delta_{\bm{R}\bm{R}'} + \delta_{\bm{R}\bm{R}''}\big) W_{\beta\bm{R}'}^*(\bm{x}) W_{\alpha\bm{R}''}(\bm{x}).
\end{equation}
Expanding the relator in Eq. (\ref{relatorS}) along a straight-line path \cite{Mahon2019} and using the first line of Eq. (\ref{dipolemoments}) for the electric dipole moment, we find for the first-order electric dipole moment
\begin{align}
    \mu_{\bm{R}}^{(1)i}(\omega) = \frac{e}{2} \sum_{\alpha\beta\bm{R}'\bm{R}''} (\delta_{\bm{R}\bm{R}'} + \delta_{\bm{R}\bm{R}''}) \left[\int \mathrm{d}\bm{y}\, W_{\beta\bm{R}'}^*(\bm{y})(y^i - R^i) W_{\alpha\bm{R}''}(\bm{y})\right]\eta_{\alpha\bm{R}'';\beta\bm{R}'}^{(1)}(\omega).
\end{align}
Inserting our expression (\ref{firstorderSPDMWannier}) for the first-order SPDM, and using the dipole-moment identity (\ref{usefulidentity1}) together with the transformation relation (\ref{BlochWannierconnection}) between the local component representations of the non-Abelian Berry connection in the Bloch and Wannier frames, this becomes
\begin{align}
    \mu_{\bm{R}}^{(1)i}(\omega) &= e^2 \Omega_{\text{uc}} E^{\ell}(\omega) \sum_{mn} f_{nm} \fint_{\text{BZ}} \frac{\mathrm{d}\bm{k}}{(2\pi)^3}\, \frac{\big(\xi_{nm}^i(\bm{k}) + \mathcal{W}_{nm}^i(\bm{k})\big)\xi_{mn}^{\ell}(\bm{k})}{E_{m\bm{k}} - E_{n\bm{k}} - \hbar(\omega + i0^+)},
\end{align}
which, after spatially averaging using the first line of Eq. (\ref{spatialaverage}) \cite{Mahon2020a}, leads to the desired expression (\ref{macroscopicpolarization2}).

To obtain the macroscopic free current, we begin with Eq. (\ref{microscopicfreecurrent}) for the microscopic free current density, together with the following expression for the link currents \cite{Mahon2019}
\begin{equation}
    I(\bm{R},\bm{R}';t) = \frac{e}{i\hbar} \sum_{\alpha\beta} \Big(\bar{H}_{\alpha\bm{R};\beta\bm{R}'}(t) \eta_{\beta\bm{R}';\alpha\bm{R}}(t) - \eta_{\alpha\bm{R};\beta\bm{R}'}(t) \bar{H}_{\beta\bm{R}';\alpha\bm{R}}(t)\Big),
\end{equation}
from which the first-order contribution to the free current density is
\begin{align}
    j_F^{(1)i}(\bm{x},\omega) &= \frac{e}{2i\hbar} \sum_{\alpha\beta\bm{R}\bm{R}'} s^i(\bm{x};\bm{R},\bm{R}') \Big(H_{\alpha\bm{R};\beta\bm{R}'}^{(0)} \eta_{\beta\bm{R}';\alpha\bm{R}}^{(1)}(\omega) - \eta_{\alpha\bm{R};\beta\bm{R}'}^{(1)}(\omega) H_{\beta\bm{R}';\alpha\bm{R}}^{(0)}\Big)\nonumber \\
    &+ \frac{e}{2i\hbar} \sum_{\alpha\beta\bm{R}\bm{R}'} s^i(\bm{x};\bm{R},\bm{R}') \Big(H_{\alpha\bm{R};\beta\bm{R}'}^{(1)}(\bm{R}_a,\omega) \eta_{\beta\bm{R}';\alpha\bm{R}}^{(0)} - \eta_{\alpha\bm{R};\beta\bm{R}'}^{(0)}H_{\beta\bm{R}';\alpha\bm{R}}^{(1)}(\bm{R}_a,\omega)\Big),
\end{align}
where the second line features the first-order hopping matrix elements (\ref{firstorderhopping}). We refer to the first line as a ``dynamical" contribution and the second line as a ``compositional" contribution \cite{Mahon2020}. Introducing a straight-line path for the relator (\ref{relatorS}), the dynamical contribution at the lattice site $\bm{R}$ can be written
\begin{equation}
    j_{F\bm{R}}^{(1.\text{I})i}(\bm{x},\omega) = \frac{e}{2\hbar i} \delta(\bm{x}-\bm{R}) \sum_{\alpha\beta\bm{R}'} (R^i-R'^i)\Big(H_{\alpha\bm{R};\beta\bm{R}'}^{(0)}\eta_{\beta\bm{R}';\alpha\bm{R}}^{(1)}(\omega) - \eta_{\alpha\bm{R};\beta\bm{R}'}^{(1)}(\omega) H_{\beta\bm{R}';\alpha\bm{R}}^{(0)}\Big).
\end{equation}
We insert the first-order SPDM (\ref{firstorderSPDMWannier}) into this expression, and, after implementing our spatial averaging procedure, we find for the dynamical modification of the macroscopic free current density
\begin{align}
    J_{F}^{(1.\text{I})i}(\omega) = \frac{ie^2}{\hbar } E^{\ell}(\omega) \sum_{mn} f_{nm}  \fint_{\text{BZ}} \frac{\mathrm{d}\bm{k}}{(2\pi)^3}\,\mathcal{W}_{nm}^i(\bm{k}) \xi_{mn}^{\ell}(\bm{k})  \left(1 + \frac{\hbar \omega}{E_{m\bm{k}} - E_{n\bm{k}} - \hbar (\omega + i0^+)}\right).
    \label{dynamicalJF}
\end{align}
The second ``compositional" contribution to the microscopic free current density is given by
\begin{equation}
    j_{F\bm{R}}^{(1.\text{II})i}(\bm{x},\omega) = \frac{e}{2\hbar i} \delta(\bm{x}-\bm{R}) \sum_{\alpha\lambda\bm{R}'} (R^i-R'^i)\Big(H_{\alpha\bm{R};\lambda\bm{R}'}^{(1)}(\bm{R}_a,\omega)\eta_{\lambda\bm{R}';\alpha\bm{R}}^{(0)} - \eta_{\alpha\bm{R};\lambda\bm{R}'}^{(0)} H_{\lambda\bm{R}';\alpha\bm{R}}^{(1)}(\bm{R}_a,\omega)\Big).
\end{equation}
Inserting the expression (\ref{unperturbedSPDM}) together with Eq. (\ref{firstorderhopping}) for the first-order hopping matrix elements, and implementing our spatial averaging procedure, we find
\begin{align}
    J_{F}^{(1.\text{II})i}(\omega) = - \frac{e^2}{\hbar} E^{\ell}(\omega) \sum_{mn} f_{nm} \fint_{\mathrm{BZ}} \frac{\mathrm{d}\bm{k}}{(2\pi)^3}\, \Im\Big[\big(\xi_{nm}^{\ell}(\bm{k}) + \mathcal{W}_{nm}^{\ell}(\bm{k})\big)\mathcal{W}_{mn}^i(\bm{k})\Big]
\end{align}
for the compositional modification of the macroscopic free current density. Combining this expression with the dynamical contribution (\ref{dynamicalJF}) yields the desired expression (\ref{macroscopicfreecurrent2}).
\end{widetext}

\section{Hall conductivity tensor}\label{AppendixD}

We derive the Hall conductivity tensor (\ref{QAHcond}) starting with the first-order macroscopic current density (\ref{J1identity}). Comparing the second term of this current density with Eq. (\ref{centralresult}), the Hall conductivity tensor is given by 
\begin{equation}
    \sigma_{\text{QAH}}^{i\ell} = \frac{e^2}{(2\pi)^3 \hbar} \varepsilon^{i \ell j} \sum_n f_n \fint_{\mathrm{BZ}} \mathrm{d}\bm{k}\, \delta_{j a} \varepsilon^{abc} \partial_b \mathcal{W}_{c,nn}(\bm{k}),
    \label{QAHintermediate1}
\end{equation}
where we have modified some of the (Cartesian) indices for consistency with the summation convention, noting that the quantity $\mathcal{W}_{nn}(\bm{k})$ involves the diagonal Bloch-frame components of the Maurer-Cartan $1$-form (\ref{Winverse}).

Suppose that the lattice $\Gamma$ is spanned by a collection of (possibly nonorthogonal) primitive lattice vectors,
\begin{equation}
    \Gamma = \mathrm{span}_{\mathbb{Z}}\big(\{\bm{a}_1,\bm{a}_2,\bm{a}_3\}\big),
\end{equation}
consituting an affine frame for this lattice. The reciprocal lattice can then be written
\begin{equation}
    \Gamma^* = \text{span}_{\mathbb{Z}}\big(\{\bm{g}_1,\bm{g}_2,\bm{g}_3\}\big),
\end{equation}
where the reciprocal lattice vectors $(\bm{g}_1,\bm{g}_2,\bm{g}_3)$ are defined through the duality relations $\bm{a}_{\alpha} \cdot \bm{g}_{\beta} = 2\pi \delta_{\alpha\beta}$ for all $1\leq \alpha,\beta \leq 3$. Being a Euclidean space, the reciprocal space $\mathbb{R}_*^d$ comes equipped with the flat Euclidean metric that reduces to the usual ``dot product" on its tangent spaces. This metric $\delta$ induces a Riemannian metric $h$ on the smooth manifold $\mathrm{BZ}$, which is defined by the pullback $\delta = Q^* h$ under the quotient map $Q : \mathbb{R}_*^d \to \mathbb{R}_*^d / \Gamma^*$ (as discussed in Appendix \ref{AppendixA}). Working in a ``Cartesian" chart for which the local coordinates are given by $(k_x, k_y, k_z)$, the chart representation of the metric $h$ is
\begin{equation}
    h = h_{ab}(\bm{k}) \mathrm{d}k^a \otimes \mathrm{d}k^b,
\end{equation}
where the Cartesian indices $a,b \in \{x,y,z\}$. The components of the metric are $h_{ab}(\bm{k}) = h(Q_*\partial_a, Q_*\partial_b)$, involving the pushforward by the quotient map $Q$ of the coordinate vector fields $\partial_a = \partial / \partial k^a$ on $\mathbb{R}_*^d$ to the $\mathrm{BZ}$. But from the flat Euclidean metric $\delta$ we have
\begin{equation}
    \delta(\partial_a, \partial_b) = (Q^* h)(\partial_a, \partial_b) = h(Q_{*}\partial_a, Q_* \partial_b),
\end{equation}
and so the metric components are $h_{ab}(\bm{k}) = \delta_{ab}$ in this Cartesian chart. With this result we can write Eq. (\ref{QAHintermediate1}) for the Hall conductivity tensor as
\begin{equation}
    \sigma_{\text{QAH}}^{i\ell} = \frac{e^2}{(2\pi)^3 \hbar} \varepsilon^{i\ell j} \sum_n f_n \fint_{\mathrm{BZ}} \mathrm{d}\bm{k}\, h_{j a}(\bm{k}) \varepsilon^{abc} \partial_b \mathcal{W}_{c,nn}(\bm{k}).
    \label{QAHintermediate2}
\end{equation}
To define the appropriate Chern numbers associated with $2$-cycles in the Brillouin zone manifold, we introduce a different chart around the point $\bm{k} \in \mathrm{BZ}$ with local coordinates $(\kappa_1, \kappa_2, \kappa_3)$. In a heuristic sense, these coordinates can be thought of as those of the point $\bm{k}$ in the ``basis" of reciprocal lattice vectors, and to differentiate them from the coordinates in the Cartesian chart, which are indexed by Latin letters, we will index them by Greek letters. In this ``reciprocal-lattice chart" the chart representation of the metric $h$ is
\begin{equation}
    h = h_{\mu\nu}(\bm{\kappa}) \mathrm{d}\kappa^{\mu}\otimes \mathrm{d}\kappa^{\nu},
\end{equation}
where the summed Greek indices $1 \leq \mu,\nu \leq 3$. Under the chart transition map from the Cartesian chart to the reciprocal-lattice chart, the components of the metric $h$ transform according to
\begin{equation}
    h_{\mu\nu}(\bm{\kappa}) = \left(\frac{\partial k^a}{\partial \kappa^{\mu}}\right) \left(\frac{\partial k^b}{\partial \kappa^{\nu}}\right) h_{ab}(\bm{k}),
    \label{hRLC}
\end{equation}
and under the inverse chart transition map we have
\begin{equation}
    h_{ab}(\bm{k}) = \left(\frac{\partial \kappa^{\mu}}{\partial k^a}\right) \left(\frac{\partial \kappa^{\nu}}{\partial k^b}\right) h_{\mu\nu}(\bm{\kappa}).
\end{equation}
To determine the form of this chart transition map, we note that the reciprocal lattice vectors for a generic crystal in three dimensions can be written in terms of the reciprocal-space unit vectors as
\begin{equation}
    \bm{g}_{\mu} = \hat{\bm{e}}_a (H^{-1})^a_{\;\; \mu},
    \label{defH}
\end{equation}
for some invertible matrix $H \in \mathrm{GL}(3,\mathbb{R})$. But if the coordinate functions $(\kappa_1, \kappa_2, \kappa_3)$ are to represent a given point $\bm{k} \in \mathrm{BZ}$ in the basis of reciprocal lattice vectors, then the transition map from the Cartesian chart to the reciprocal-lattice chart and its inverse transition map can be written, respectively, as 
\begin{align}
    \kappa^{\mu}(k_x,k_y,k_z) &= H^{\mu}_{\;\; a} k^a,\nonumber \\
    k^a(\kappa_1,\kappa_2, \kappa_3) &= (H^{-1})^a_{\;\;\mu} \kappa^{\mu},
    \label{transitionmaps}
\end{align}
for all $1 \leq \mu \leq 3$ and $a \in \{x,y,z\}$. From these expressions and Eq. (\ref{hRLC}) for the components of the metric $h$ in the reciprocal-lattice chart, we have
\begin{equation}
    h_{\mu\nu}(\bm{\kappa}) = (H^{-1})^a_{\;\;\mu} (H^{-1})^b_{\;\;\nu} \delta_{ab}.
\end{equation}
Now under the chart transition maps (\ref{transitionmaps}) the Levi-Civita symbol transforms as a tensor \textit{density}, namely
\begin{equation}
    \varepsilon^{abc} = \frac{1}{\sqrt{\det(h_{\mu\nu})}} \left(\frac{\partial k^a}{\partial \kappa^{\alpha}}\right) \left(\frac{\partial k^b}{\partial \kappa^{\beta}}\right)\left(\frac{\partial k^c}{\partial \kappa^{\gamma}}\right) \varepsilon^{\alpha\beta\gamma},
\end{equation}
\\
and so the quantities in the integrand of Eq. (\ref{QAHintermediate2}) transform according to
\begin{equation}
    \varepsilon^{abc} \partial_b \mathcal{W}_{c,nn}(\bm{k}) = \frac{1}{\sqrt{\det(h_{\mu\nu})}} \left(\frac{\partial k^a}{\partial \kappa^{\alpha}}\right) \varepsilon^{\alpha\beta\gamma} \partial_{\beta} \mathcal{W}_{\gamma,nn}(\bm{\kappa}).
\end{equation}
Also, the Riemannian volume form associated with the induced metric $h$ on the $\mathrm{BZ}$ is a tensor \textit{density} and can be written in the reciprocal-lattice chart as
\begin{equation}
    \mathrm{d}\bm{k} = \mathrm{d}\bm{\kappa}\, \sqrt{\det(h_{\mu\nu})},
\end{equation}
where we have defined
\begin{equation}
    \mathrm{d}\bm{\kappa} \equiv \mathrm{d}\kappa^1 \wedge \mathrm{d}\kappa^2 \wedge \mathrm{d}\kappa^3,
\end{equation}
from which follows
\begin{widetext}
\begin{equation}
    \sigma_{\text{QAH}}^{i\ell} = \frac{e^2}{(2\pi)^3 \hbar} \varepsilon^{i \ell j}\delta_{j a} (H^{-1})^a_{\;\;\alpha} \sum_n f_n \fint_{\mathrm{BZ}} \mathrm{d}\bm{\kappa}\, \varepsilon^{\alpha\beta\gamma} \partial_{\beta} \mathcal{W}_{\gamma,nn}(\bm{\kappa}).
\end{equation}
Being written in the reciprocal-lattice chart, we are now able to process this integral over the three fundamental $2$-cycles in the smooth manifold $\mathrm{BZ} \cong \mathbb{T}^3$. Indeed, for each $1 \leq \alpha \leq 3$ we can write the corresponding (adapted) integral in the form
\begin{align}
    \frac{1}{4\pi^2} \fint_{\mathrm{BZ}} \mathrm{d}\bm{\kappa} \sum_n f_n\, \varepsilon^{1\beta\gamma}\partial_{\beta} \mathcal{W}_{\gamma,nn}(\bm{\kappa}) &= \frac{1}{2\pi} \int_{\mathrm{BZ}} \mathrm{d}\kappa^1 \wedge \left(\frac{1}{2\pi} \sum_n f_n\, \varepsilon^{1 \beta\gamma}\partial_{\beta} \mathcal{W}_{\gamma,nn}(\bm{\kappa})\, \mathrm{d}\kappa^2 \wedge \mathrm{d}\kappa^3\right),\nonumber \\
    \frac{1}{4\pi^2} \fint_{\mathrm{BZ}} \mathrm{d}\bm{\kappa} \sum_n f_n\, \varepsilon^{2\beta\gamma}\partial_{\beta} \mathcal{W}_{\gamma,nn}(\bm{\kappa}) &= \frac{1}{2\pi} \int_{\mathrm{BZ}} \mathrm{d}\kappa^2 \wedge \left(\frac{1}{2\pi} \sum_n f_n\, \varepsilon^{2 \beta\gamma}\partial_{\beta} \mathcal{W}_{\gamma,nn}(\bm{\kappa})\, \mathrm{d}\kappa^3 \wedge \mathrm{d}\kappa^1\right),\nonumber \\
    \frac{1}{4\pi^2} \fint_{\mathrm{BZ}} \mathrm{d}\bm{\kappa} \sum_n f_n\, \varepsilon^{3\beta\gamma}\partial_{\beta} \mathcal{W}_{\gamma,nn}(\bm{\kappa}) &= \frac{1}{2\pi} \int_{\mathrm{BZ}} \mathrm{d}\kappa^3 \wedge \left(\frac{1}{2\pi} \sum_n f_n\, \varepsilon^{3 \beta\gamma}\partial_{\beta} \mathcal{W}_{\gamma,nn}(\bm{\kappa})\, \mathrm{d}\kappa^1 \wedge \mathrm{d}\kappa^2\right).
\end{align}
\end{widetext}
Having included the trace over occupied valence bands in the integrand on the right-hand-side, we can now drop the adapted integral notion, because, as we have shown previously \cite{CImanuscript1}, any contributions coming from band crossings between two intersecting bands will be equal in magnitude and opposite in sign, leading to cancellations for all points in the locus of degeneracies $\mathscr{D}_{\mathcal{B}}$ across the Brillouin zone. To proceed, we introduce a triple of smooth embedding maps $\phi_{s}^{(\alpha)}: \mathbb{T}^2 \hookrightarrow \mathrm{BZ}$ for all $1\leq \alpha\leq 3$ defined by
\begin{align}
    \phi_{s}^{(1)}(\kappa_2,\kappa_3) = (s, \kappa_2,\kappa_3),\nonumber \\
    \phi_{s}^{(2)}(\kappa_1,\kappa_3) = (\kappa_1, s,\kappa_3),\nonumber \\
    \phi_{s}^{(3)}(\kappa_1,\kappa_2) = (\kappa_1, \kappa_2,s),
\end{align}
in which case the $\alpha\text{th}$ $2$-cycle in the $\mathrm{BZ}$ is then the image set $\mathbb{T}^2(s) = \phi_{s}^{(\alpha)}(\mathbb{T}^2)$ with $s \in \mathbb{S}^1$ being fixed. Focusing on the case $\alpha = 1$, it is straightforward to show that \cite{CImanuscript1}
\begin{align}
    \frac{1}{2\pi} \sum_n f_n\, \varepsilon^{1 \beta\gamma}\partial_{\beta} \mathcal{W}_{\gamma,nn}(\bm{\kappa})\, \mathrm{d}\kappa^2 \wedge \mathrm{d}\kappa^3\nonumber \\
    = \big(\phi_{s}^{(1)}\big)^*\left(\frac{1}{2\pi} \sum_n f_n \mathrm{d}\mathcal{W}_{nn}(\bm{\kappa})\right).
\end{align}
Now if there exists a Wannier frame for the Bloch bundle from which we can construct ELWFs --- this is equivalent to the Bloch bundle being trivializable as a \textit{holomorphic} Hilbert bundle \cite{Brouder, Freed} --- then one can show  \cite{CImanuscript1}
\begin{equation}
    \sum_n f_n \Big(\mathrm{d}\xi_{nn}(\bm{\kappa}) + \mathrm{d}\mathcal{W}_{nn}(\bm{\kappa})\Big) = 0,
\end{equation}
from which follows
\begin{align}
    \frac{1}{2\pi} \sum_n f_n\, \varepsilon^{1 \beta\gamma}\partial_{\beta} \mathcal{W}_{\gamma,nn}(\bm{\kappa})\, \mathrm{d}\kappa^2 \wedge \mathrm{d}\kappa^3\nonumber \\
    = - \big(\phi_{s}^{(1)}\big)^*\left(\frac{1}{2\pi} \sum_n f_n \mathrm{d}\xi_{nn}(\bm{\kappa})\right).
\end{align}
The quantity in parentheses is the first Chern character of the valence bundle \cite{NakaharaBook}, given by 
\begin{equation}
    \mathrm{ch}_1(F_{\mathcal{V}}(\bm{\kappa})) = \frac{1}{2\pi} \sum_n f_n \mathrm{d}\xi_{nn}(\bm{\kappa}),
    \label{Cherncharacter}
\end{equation}
where $F_{\mathcal{V}}$ is the (Berry) curvature $2$-form of which the Cartesian components are given by Eq. (\ref{Berrycurvaturecomponents}). Following the same reasoning for $\alpha = 2,3$, we have
\begin{widetext}
\begin{align}
    \frac{1}{4\pi^2} \fint_{\mathrm{BZ}} \mathrm{d}\bm{\kappa} \sum_n f_n\, \varepsilon^{\alpha\beta\gamma}\partial_{\beta} \mathcal{W}_{\gamma,nn}(\bm{\kappa}) = - \frac{1}{2\pi} \int_{\mathrm{BZ}} \mathrm{d}s \wedge \Big(\big(\phi_{s}^{(\alpha)}\big)^*\big(\mathrm{ch}_1(F_{\mathcal{V}}(\bm{\kappa}))\big)\Big).
    \label{int1}
\end{align}
We use the ``co-area formula" to simplify this expression, which is a generalization of Fubini's theorem to smooth and Riemannian manifolds \cite{LeeBook}. By application of this formula to Eq. (\ref{int1}) we find
\begin{equation}
    - \frac{1}{2\pi} \int_{\mathrm{BZ}} \mathrm{d}s \wedge \Big(\big(\phi_{s}^{(\alpha)}\big)^*\big(\mathrm{ch}_1(F_{\mathcal{V}}(\bm{\kappa}))\big)\Big) = - \frac{1}{2\pi} \int_{\mathbb{S}^1} \mathrm{d}s \left(\frac{1}{2\pi} \int_{\mathbb{T}^2(s)} \big(\phi_{s}^{(\alpha)}\big)^*\big(\text{ch}_1(F_{\mathcal{V}}(\bm{\kappa}))\big)\right).
\end{equation}
\end{widetext}
For each fixed $s \in \mathbb{S}^1$ the right-hand-side features the integral of the first Chern character (\ref{Cherncharacter}) over a closed $2$-dimensional submanifold of the $\mathrm{BZ}$, which is guaranteed to be an integer by a fundamental result of Chern-Weil theory \cite{NakaharaBook}. This is precisely the first Chern number of the valence bundle $\mathcal{V}$ associated with this $2$-cycle, namely
\begin{equation}
    C_{\mathcal{V}}^{\alpha}(s) = \frac{1}{2\pi} \int_{\mathbb{T}^2(s)} \big(\phi_{s}^{(\alpha)}\big)^*\big(\text{ch}_1(F_{\mathcal{V}}(\bm{\kappa}))\big) \in \mathbb{Z},
\end{equation}
from which follows
\begin{align}
    \frac{1}{4\pi^2} \fint_{\mathrm{BZ}} \mathrm{d}\bm{\kappa} \sum_n f_n\, \varepsilon^{\alpha\beta\gamma}\partial_{\beta} \mathcal{W}_{\gamma,nn}(\bm{\kappa}) = - \frac{1}{2\pi} \int_{\mathbb{S}^1} \mathrm{d}s\, C_{\mathcal{V}}^{\alpha}(s).
\end{align}
But the integrand on the right-hand-side of this expression is an integer-valued continuous function and is therefore constant \cite{VanderbiltBook}. Parametrizing the circle $\mathbb{S}^1$ by $s \in [0,2\pi]$ under the identification $0 \sim 2\pi$ leads to
\begin{align}
    \frac{1}{4\pi^2} \fint_{\mathrm{BZ}} \mathrm{d}\bm{\kappa} \sum_n f_n\, \varepsilon^{\alpha\beta\gamma}\partial_{\beta} \mathcal{W}_{\gamma,nn}(\bm{\kappa}) = - C_{\mathcal{V}}^{\alpha},
\end{align}
and so the Hall conductivity tensor is 
\begin{equation}
    \sigma_{\mathrm{QAH}}^{i\ell} = \frac{e^2}{2\pi \hbar} \varepsilon^{i j \ell} \delta_{j a} (H^{-1})^{a}_{\;\;\alpha} C_{\mathcal{V}}^{\alpha}.
\end{equation}
Using Eq. (\ref{defH}) for the chart transition map, we find
\begin{equation}
    \delta_{j a} (H^{-1})^a_{\;\;\alpha} = \hat{\bm{e}}_{j} \cdot \bm{g}_{\alpha},
\end{equation}
and so the Hall conductivity tensor for the quantum anomalous Hall effect is given by Eq. (\ref{QAHcond}), as desired.

\bibliographystyle{apsrev4-1}
\bibliography{ChernInsulator.bib}

\end{document}